# BARCODE: Biomaterial Activity Readouts to Categorize, Optimize, Design and Engineer for high throughput screening and characterization of dynamically restructuring soft materials

Qiaopeng Chen[†1], Aditya Sriram[†2], Ayan Das[1], Katarina Matic[2], Maya Hendija[2], Keegan Tonry[3], Jennifer L. Ross[4], Moumita Das[3], Ryan J. McGorty[2], Rae M. Robertson-Anderson[2*], Megan T. Valentine[1*]

[1] Department of Mechanical Engineering, University of California, Santa Barbara
[2] Department of Physics and Biophysics, University of San Diego
[3] School of Physics and Astronomy, Rochester Institute of Technology
[4] Department of Physics, Syracuse University

[*] Corresponding authors, equal contribution
[†] Equal contribution

## ABSTRACT

Active, responsive, non-equilibrium materials–at the forefront of materials engineering–offer dynamical restructuring, mobility and other complex life-like properties. Yet, this enhanced functionality comes with significant amplification of the size and complexity of the datasets needed to characterize their properties, thereby challenging conventional approaches to analysis. To meet this need, we present BARCODE: Biomaterial Activity Readouts to Categorize, Optimize, Design and Engineer, an open-access software that automates high throughput screening of microscopy video data to enable non-equilibrium material optimization and discovery. BARCODE produces a unique fingerprint or 'barcode' of performance metrics that visually and quantitatively encodes dynamic material properties with minimal file size. Using three complementary material-agnostic analysis branches, BARCODE significantly reduces data dimensionality and size, while providing rich, multiparametric outputs and rapid tractable characterization of activity and structure. We analyze a series of datasets of cytoskeleton networks and cell monolayers to demonstrate BARCODE's abilities to accelerate and streamline screening and analysis, reveal unexpected correlations and emergence, and enable broad non-expert data access, comparison, and sharing.

# INTRODUCTION

Biological activity manifests through a wide range of mechanical and dynamical features, such as stiffening, restructuring and flow. Examples include migrating slime molds, developing tissues, and the cytoskeleton, a protein network in living cells that undergoes dramatic remodeling during cell division, crawling, and wound healing [1-7]. Engineered materials can recapitulate many such features, and offer stimuli-responsiveness, patterning, and spatiotemporal control that could, in principle, be harnessed for applications such as self-healing infrastructure, dynamic prosthetics, and self-sensing protective barriers [8-19]. Like their biological analogs, engineered materials exhibiting such functionalities must be multicomponent and multiphase, with structures and dynamics operating over a broad range of length and time scales. For example, in vitro networks that recapitulate properties of the cell cytoskeleton, such as contractility, self-organization, and responsivity to external stimuli, comprise filamentous proteins, such as actin and microtubules, enzymatically-active motor proteins, such as myosin and kinesin, and a host of crosslinking proteins [12,20-30]. Varying the network formulation and intermolecular interactions can tune the material dynamics and structure over decades of spatiotemporal scales [20,25,29,31-33]. Similarly, cells in vitro exhibit dynamics and restructuring similar to living tissues such as jamming, flow, and collective patterns emerging on length scales much longer than the individual agents driving the motion [34-39]. This dynamic structural heterogeneity, foundational to function, challenges classic approaches to material design, characterization and deployment.

Complex active dynamics and restructuring presents several challenges to developing predictive relationships between material formulation and performance, and to realizing tractable engineering designs. Video sizes and complexity limit data sharing and use of standard software to process images within manageable time frames, often forcing valuable information to be ignored, discarded, or siloed. Active materials require a higher dimensionality of characterization metrics, which often emerge in unexpected ways and on spatiotemporal scales that are difficult to predict a priori. Inconsistencies in metrics, definitions, and approaches further hinder the identification of performance intersections among materials. These complexities demand readily accessible material-agnostic algorithms that enable rapid screening of large datasets and characterization of emergent dynamics in a manner that enables data- and physics-driven modeling and rational material design.

To address these challenges, we present BARCODE: Biomaterial Activity Readouts to Categorize, Optimize, Design and Engineer, an open-access software to facilitate the democratized discovery and optimization of non-equilibrium materials. BARCODE automates high throughput (HTP) screening of optical microscopy videos and outputs a unique fingerprint that encodes dynamic material properties (Fig. 1). Consisting of three complementary branches that leverage standardized image analysis approaches (Fig. 1D), BARCODE produces a unique array or 'barcode' of performance metrics for each video, and the collective dataset (Fig. 1E), significantly reducing data dimensionality, complexity and size, while providing rich, multiparametric outputs. Importantly, screening is performed without consideration of material composition or formulation, allowing unexpected correlations between performance metrics or disparate material systems to be revealed. To produce each barcode, the software calculates rich reduced data structures (RDS) and archives them to enable subsequent hypothesis-driven research. Through these features, BARCODE not only streamlines on-system screening and accelerates analysis, but also enables non-expert data access and sharing across different materials and communities.

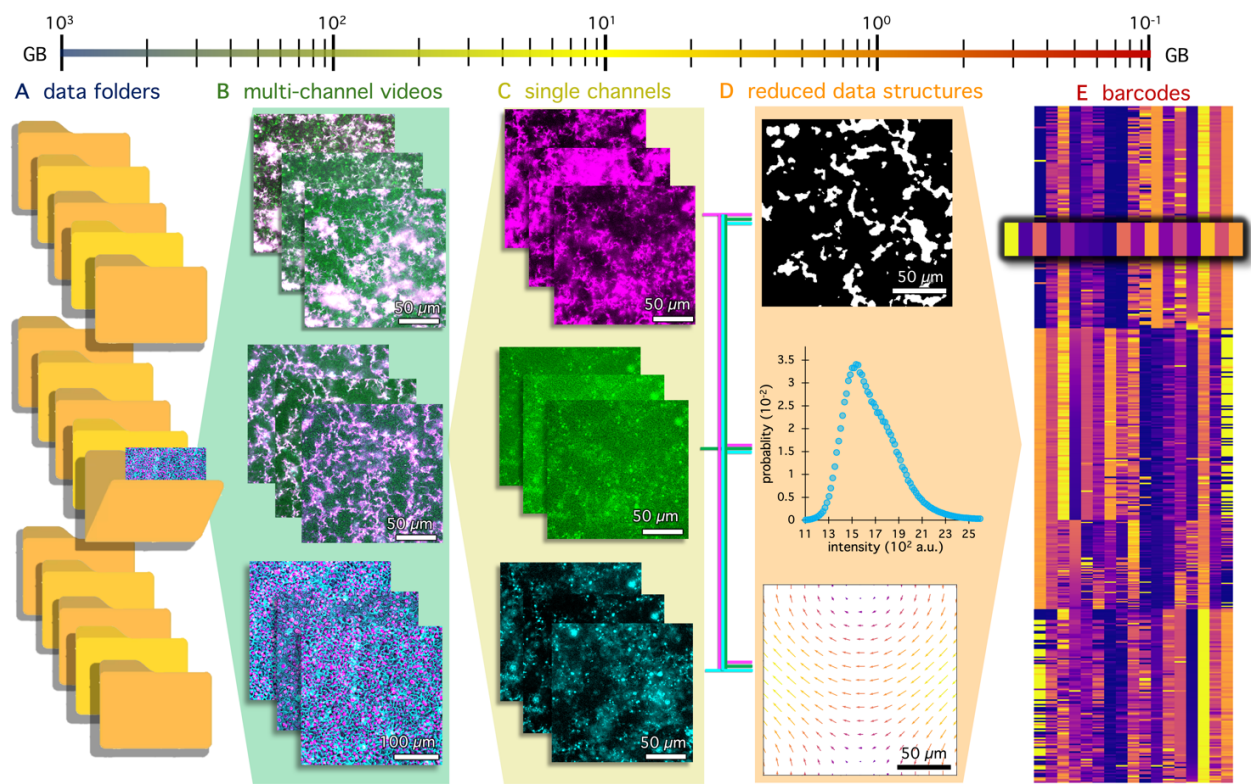

**Figure 1. BARCODE performs HTP screening of large video-based datasets to extract an information-rich, sparse-data barcode that reduces dataset size by 4 orders of magnitude. (A)** In typical experiments, a large number of multi-channel microscopy videos are acquired and saved in a hierarchical folder structure with **(B)** individual multi-channel videos within each folder, representing various compositions of active materials. Shown from top to bottom are kinesin-driven composites of actin (green) and microtubules (magenta) with (top) and without (middle) passive crosslinkers; and (bottom) a monolayer of cells with separately labeled nuclei (magenta) and cytoplasm (cyan). **(C)** Each multichannel video is separated into stacks of single-channel images. Shown from top to bottom are microtubules, actin, and microtubule crosslinkers (ASE1, anaphase spindle elongation protein 1) separated from the merged images shown in B (top). **(D)** Each channel is fed into three independent analysis branches that leverage vetted image analysis methods: binarization (top), pixel intensity distributions (middle), and optical flow (bottom). The reduced data structures (RDS) that each branch computes are saved for further screening or hypothesis-driven research outside of BARCODE; and used to extract a low dimensional 1×17 matrix or 'barcode' which visually and quantitatively represents key output metrics characterizing material structure, reconfiguration and dynamics. **(E)** Each video channel is reduced to a unique barcode that is compiled into a consolidated array providing a comprehensive and quantitative fingerprint of the dataset. The different colors depict the values of each metric from the minimum (blue) to maximum (yellow) assigned values using the plasma colorscale.

## RESULTS

## BARCODE: Biomaterial Activity Readouts to Categorize, Optimize, Design and Engineer

BARCODE is designed to enable rapid HTP analysis of multi-channel microscopy videos to screen for desirable features of active material systems (Fig. 1). By extracting key parameters that describe material structure and dynamics, BARCODE reduces large (~1 TB) and complex datasets (Fig. 1A-C) by up to four orders of magnitude. BARCODE's core architecture executes three complementary, yet independent, 'branches' in parallel (Fig. 1D). We designed each branch, the algorithmic details of which are fully described in SI Section 1 and depicted in Figure 2, to leverage established image processing tools–(1) binarization, (2) pixel intensity distribution analysis, and

(3) optical flow[40]. Each branch produces distinct metrics selected to encode key structural and dynamic features of active materials, organized in a 1×17 'barcode' that is output both numerically and as a color-coded array (Fig. 1E). We compute metrics using conventional reduced data structures (RDS) (Fig. 1D) – (1) binarized video images, (2) pixel intensity distributions, and (3) velocity fields – which are automatically archived to facilitate future downstream analysis. Importantly, BARCODE is material-agnostic, modular and highly adaptable: branches and metrics can be easily added or removed without impacting other metrics, thus providing flexibility, while allowing for discovery of unexpected correlations or trends.

To demonstrate the BARCODE workflow, we analyze two representative videos of motor-driven actin-microtubule composites displaying distinct structures and dynamics (Fig. 2A). The image Binarization (IB) branch (Fig. 2B) converts each grayscale image of the video into a binary image of white (1) and black (0) pixels using a user-defined threshold (Fig. 2B, top). From each binarized image, which is a saved RDS, we calculate the relative areas of connected regions of white pixels, 'islands', and black pixels, 'voids', and define 'connectivity' of 1 (or 0) for images having a continuous path (or not) of white pixels extending from edge to edge. We identify the maximum island area $I_i$, maximum void area $V_i$ and connectivity $C_i$ in each frame $i$, and evaluate how they change over time to determine the global maximum island ($I$) and void ($V$) area, their change from the first to last frame ($\Delta I, \Delta V$), and the fraction of frames that are connected $C = \langle C_i \rangle$ where the angled brackets indicate a time average (Fig. 2B bottom). To characterize the initial structure, we compute the maximum and secondary maximum island areas in the first frame of each video $I_{0,1}$ and $I_{0,2}$. The 7 metrics of the IB branch (Fig. 2E) accurately describe the time-varying structures shown in Fig. 2A. Video 1 remains connected and undergoes minimal restructuring, while Video 2 rapidly loses connectivity and shows more dramatic changes to island and void areas.

The Intensity Distribution (ID) branch (Fig. 2C) evaluates the probability distribution of pixel intensities, which serves as a proxy for mass density, and outputs 6 barcode entries that describe the distribution features and how they change over the course of the video to characterize the time-varying structure. We evaluate distributions of the initial and final 10% of frames in each video, which are saved as RDS, to provide a rapid information-rich assessment. In Video 1, we observe that the mode and mean intensity shift to lower values in the final frame compared to the initial, and the low-intensity mode exhibits a more pronounced peak, both attributed to the formation of aggregates (brighter regions) and more network-sparse (darker) regions. The initial and final distributions for Video 2 capture the more substantial reconfiguration compared to Video 1, with the final distribution displaying substantially more high-intensity and saturated pixels and a modest low-intensity peak, both absent in the initial distribution. These features reflect the formation of a large amorphous aggregate in the corner which moves further into the field of view over the course of the video (Fig. 2A, 6 min).

To quantitatively characterize and reduce the dimensionality of the distributions and their changes over time, BARCODE computes the median skewness $S_{i,1}$, mode skewness $S_{i,2}$, and kurtosis $K_i$ for each evaluated frame $i$ (Fig. 2C) and reports the maximum of each across all evaluated frames as $S_1$, $S_2$ and $K$ (Fig. 2E). We also evaluate distinct maxima for the first and last 10% of frames and compute their differences: $\Delta S_1$, $\Delta S_2$ and $\Delta K$ to determine the degree and type of restructuring. For example, more positive mode skewness may correlate with bundling and aggregation as it shifts the distribution to having a more pronounced low-intensity mode and high-intensity 'tail'. Positive (or negative) kurtosis values indicate less (or more) uniform distributions, which can signify material coarsening (or homogenization).

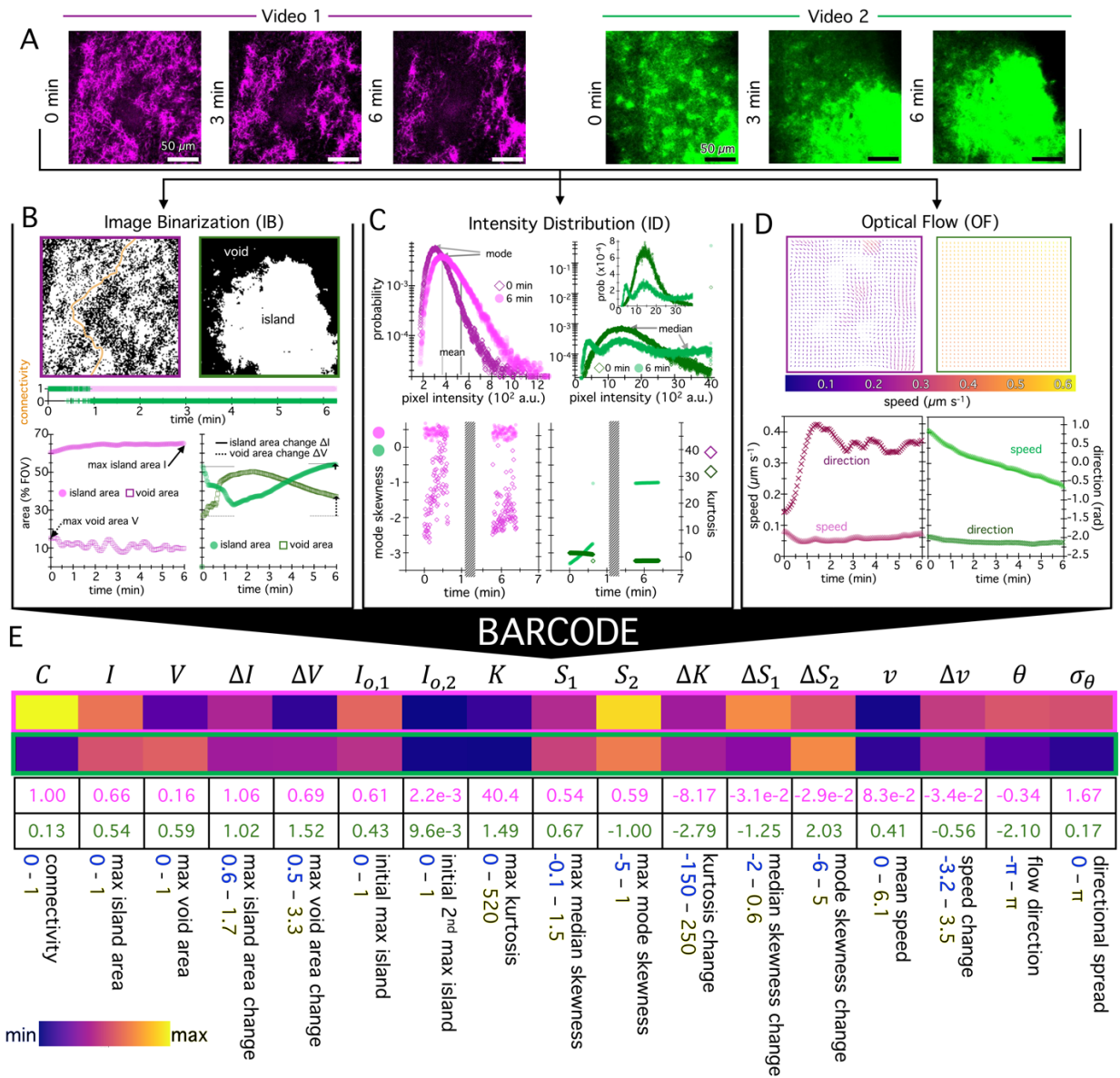

**Figure 2. BARCODE workflow leverages three independent analysis branches to rapidly extract a structural and dynamical fingerprint. (A)** Confocal fluorescence microscopy videos of two independent fields of view (FOV) show microtubules (magenta, Video 1) and actin (green, Video 2) undergoing motor-driven restructuring, shown by images acquired at the beginning (0 min), middle (3 min) and end (6 min) of the two experiments. Video 1 demonstrates local contraction, characterized by several bundling points within the FOV. Video 2 illustrates large-scale contraction. Each video is analyzed by three independent branches **(B-D)**. **(B)** The Image Binarization (IB) branch binarizes each image of each video via a user-defined threshold, resulting in a down-sampled binary video RDS. Shown are the 0 min frame of Video 1 (magenta border) and 6 min frame of Video 2 (green border), with full connectivity (orange line demonstrates a fully connected path with line width exaggerated for ease of visualization), and largest island and void (contiguous white and black region), respectively, depicted. Plot of connectivity vs time, shown below for Video 1 (magenta) and Video 2 (green), is used to compute the fractional connectivity metric $C$. Plots of the maximum island and void areas in each frame versus time are used to determine BARCODE metrics: global maximum island and void areas, $I$ and $V$, and their changes between the beginning to end of the video, $\Delta I$ and $\Delta V$. Initial maximum and 2nd maximum island areas $I_{0,1}$ and $I_{0,2}$ are also determined in this branch. **(C)** The Intensity

Distribution (ID) branch evaluates the distribution of pixel intensities for each image in the first and last 10% of frames. The top plots display the RDS distributions for the first and last frames of Video 1 (magenta) and Video 2 (green) with key statistical markers, including mode, mean and median, indicated, and showing the change in the distributions over time. Video 2 inset shows each distribution on linear y-scale with the highest intensity (saturated) point excluded. Bottom plots show the mode skewness (left axis, filled circles) and kurtosis (right axis, open diamonds) vs time for the initial and final 10% of frames of the two example videos, which are used to determine maximum kurtosis $K$, maximum mode skewness $S_2$ and their changes from the initial to final 10% of frames, $\Delta K, \Delta S_2$. The ID branch also computes the maximum median skewness $S_1$ and its change $\Delta S_1$. **(D)** The Optical Flow (OF) branch generates a velocity field for each image of the video downsampled by a user-defined interval (default 8×8 pixel downsampling shown). The first flow field of this RDS is shown for Videos 1 (left) and 2 (right), with the arrow length and color indicating speed relative to individual flow fields and the group (see colorscale), respectively, and direction indicating local material velocity direction. Bottom plots show corresponding speed and flow direction versus time, evaluated to extract BARCODE metrics: mean speed $v$, speed change $\Delta v$, flow direction $\theta$, and the standard deviation of $\theta$ across frames $\sigma_\theta$. **(E)** The lowest dimensional BARCODE output: a 1×17 matrix of key quantitative metrics that are displayed visually using a standard color scale to enable rapid comparison and pattern identification of numerical metric values. Top row: Video 1 (magenta border); bottom row: Video 2 (green border). Each cell within the barcode is labeled by the metric variable (top labels), as well as its name and corresponding minimum (blue outlined text) and maximum (yellows outlined text) value of the colorscale (bottom labels). Table 1 provides definitions of each metric and full algorithmic details describing the calculation of each metric can be found in the Supplemental Information.

The Optical Flow (OF) branch calculates velocity vector fields for a subset of evenly spaced frames with a user-defined interval (Fig. 2D), which we save as RDS, and evaluates them to determine 4 barcode metrics that quantify the material dynamics. Specifically, the OF branch computes the mean speed $v$ and flow direction $\theta$, each averaged across all vectors of all frames, as well as the change in speed between the final and initial vector fields $\Delta v$, and directional spread $\sigma_\theta$, calculated as the standard deviation of flow direction across all vectors of all fields. Together these metrics determine the magnitude and direction of the material motion, and the extent to which it is directed or randomly oriented, and speeding up or slowing down. OF analysis of the two example videos reveal starkly different dynamics (Fig. 2D). Video 1 flow field has smaller, less variable, and more randomly oriented vectors whereas Video 2 has a higher, more directed speed that steadily decreases over time, reflected in the different colors of the last 4 entries of their respective barcodes (Fig. 2E).

BARCODE can be executed on dozens, even hundreds of videos to produce a single barcode array (Fig. 1D), in 1-2 minutes per GB (SI Table S1), that can be used to identify patterns, correlations and trends. To benchmark BARCODE performance and demonstrate its ability to rapidly screen and discover properties of a broad range of non-equilibrium materials, we present the results of BARCODE analyses performed on microscopy data for four different materials that each have different image sizes, resolution, durations, acquisition parameters, and dataset sizes.

**BARCODE accurately measures filament speeds and reveals emergent correlations in active cytoskeleton composites**

We first analyze a published set of two-channel confocal videos of composites of entangled actin and microtubules, with and without crosslinkers, that undergo restructuring via kinesin and myosin motors acting on microtubules and actin, respectively (Fig. 3A) [20]. The dataset includes 48 videos of ~1000 frames of size 512×512, each with two channels that separately visualize the actin and microtubules (Fig. 3B). Originally, the videos were analyzed using differential dynamic microscopy (DDM) [41,42], an established and powerful, yet labor-intensive, approach to show that all formulations exhibit ballistic motion with speeds spanning roughly three orders of magnitude. DDM analysis also revealed that the data could be categorized into three distinct dynamic classes: 'fast' directed flow, 'slow' isotropic restructuring or 'multi-mode' dynamics that manifested aspects of fast and slow behavior (Fig. 3B). Importantly, the dynamic class was not statistically correlated with the material formulation and the molecular underpinnings remained unclear.

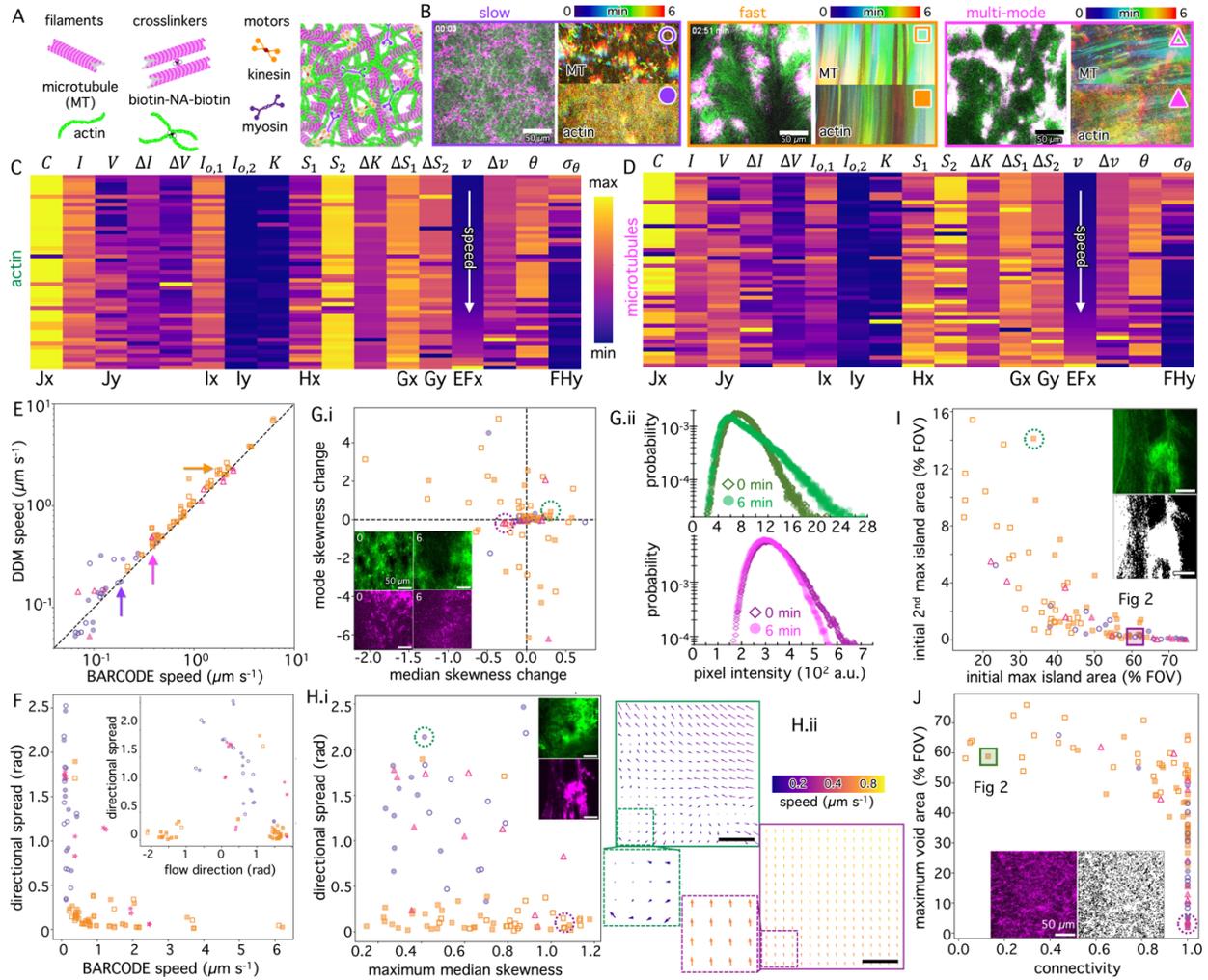

**Figure 3: BARCODE validation and discovery of emergent properties and relationships in motor-driven cytoskeletal composites.** **(A)** Schematic of the components of the cytoskeleton composite[20] including actin (green), microtubules (MTs, magenta), passive crosslinkers (biotin-NeutrAvidin-biotin), and multimeric myosin (dark purple) and kinesin (orange) motors. **(B)** Multi-channel images of actin (green) and microtubules (magenta) and temporal colormaps for each channel showing network motion during representative 6-min videos and labelled according to the three previously identified dynamical classes (slow: purple circles, fast: orange squares, multi-mode: magenta triangles), and filament types (actin: filled, microtubules:open), as used in plots E-J. **(C,D)** Barcode matrix for the actin **(C)** and microtubule **(D)** channels of the source videos [20] ordered by increasing mean speed. Letters below select columns denote the figure subpanel **(E-J)** and axis (x,y) where the metric is plotted. **(E)** Mean speed $v$ calculated with BARCODE plotted against speeds calculated using differential dynamic microscopy (DDM)[20], show excellent agreement over two orders of magnitude (p-values < 0.01 in all cases, see SI Table S2). The dashed equivalency line indicates where the two values are equal. **(F-J)** Correlations between different BARCODE metrics validate prior findings [20] and reveal new relationships and behaviors. **(F)** Correlation between mean speed $v$ and directional spread $\sigma_\theta$ shows faster speeds are correlated with directed motion (minimal $\sigma_\theta$). (Inset) Correlation between $\sigma_\theta$ and flow

direction $\theta$ shows fast flows are directed along $\pm\pi$ while slow and multi-mode videos have no preferred direction. **(G)** Correlation between median skewness change and mode skewness change with dashed lines denoting no change (i). (inset) The initial (0) and final (6) frames and (ii) intensity probability distribution RDS for the data points enclosed by color-matched dashed circles. The green circled data shows more pronounced changes between the initial and final probability distributions and images compared to the magenta circled data. The slight negative values of the latter reflect the slight shrinking of the high-intensity distribution tail, compared to its large extension for the former. **(H)** Correlation between flow directional spread (also plotted in F) and maximum median skewness. The frames (inset) and down-sampled (16×16 pixel) flow field RDS (ii) at which the maximum median skewness is reached for the data points enclosed by color-matched dashed circles, which show more randomly oriented motion coupled with more uniform structure (green) and more directed motion coupled with more anisotropic structure (magenta). Intensity distributions and full resolution (8×8 down-sampling) flow fields for the circled data are shown in SI Figs. S1, S2. **(I)** Correlation between initial maximum island area $I_{0,1}$ and initial secondary maximum island area $I_{0,2}$ show slow and multi-mode composites are initially dominated by a single large island while fast composites have multiple islands with a broader distribution of initial sizes. Inset shows raw image and binarized RDS for the data point encircled in green. The boxed-in magenta data point corresponds to Video 1 (Fig. 2). **(J)** Correlation between connectivity and maximum void area show slow and multi-mode networks are largely connected with the maximum void comprising <50% of the FOV. Fast data is often not connected with the largest void filling most of the FOV. Inset images are the raw image and binarized RDS for the data point encircled in magenta. The boxed-in green data point corresponds to Video 2 (Fig. 2). Scale bars are 50 μm for all images.

BARCODE analysis of this dataset took ~9 mins to complete, with no pre- or post-processing of the videos, compared to ~53 min required to complete the first step of DDM analysis, after which several rounds of subjective fitting and parameter choices requiring expertise are needed to determine dynamics. The resulting barcode matrices for the actin (Fig. 3C) and microtubule (Fig. 3D) channels also provide structural and orientational data for the complete dataset, which was not comprehensively analyzed in the original work[20]. The 4 OF entries are indistinguishable between the two channels, consistent with the nontrivial finding that actin and microtubules exhibited similar speeds across the entire formulation space[20]. The other branches show differences, indicating distinct and previously unreported, structural features of actin and microtubule networks.

To first benchmark and validate BARCODE, we compare the mean speeds $v$ computed from BARCODE to the previously reported DDM-computed speeds for the entire dataset, color-coded by the nominal dynamical class (Fig. 3B,E). We find remarkably good agreement over two decades of speeds (Fig. 3E, SI Table S2). Multi-mode class data deviates from the equivalency line more than the other classes because the original DDM analysis resulted in two speeds, which were averaged here, while BARCODE computes a single speed for each condition. BARCODE also quantitatively captures other dynamical features of each class that were only qualitatively described in prior work. Specifically, prior particle image velocimetry (PIV) analysis demonstrated that a single exemplar video for each of the fast and slow classes exhibited unidirectional and randomly oriented motion, respectively. By evaluating correlations between mean speed $v$, directional spread $\sigma_\theta$, and flow direction $\theta$, BARCODE not only reproduces this result but demonstrates its applicability to all data in each class (Fig. 3F).

We next demonstrate the ability of BARCODE to rapidly discover structural properties and correlations. We observe substantially greater median and mode skewness changes, $\Delta S_1$ and $\Delta S_2$, for fast class data compared to slow and multi-mode (Fig. 3Gi), suggesting more pronounced restructuring. Moreover, the larger range in $\Delta S_2$ values compared to $\Delta S_1$ is a likely indicator of pixel saturation due to largescale aggregation, shifting the mode from a low intensity peak to the highest value (as shown in Fig. 2C). Focusing on two example data points with universally increased skewness ($\Delta S_1, \Delta S_2 > 0$, green circle) and decreased skewness ($\Delta S_1, \Delta S_2 < 0$, magenta circle) (Fig. 3Gi), we find that the distributions and images support the BARCODE metrics. Namely, the distributions (Fig. 3G.ii) show increased (green) or decreased (magenta) high intensity tails and left or right shifted peaks, correlating with increased or decreased skewness, reflective of increased or decreased brightness and heterogeneity in the images (Fig. 3Gi inset).

A powerful feature of BARCODE is its ability to directly correlate structural and dynamical features, such as the relationship between directional spread $\sigma_\theta$ and maximum median skewness $S_1$ (Fig. 3Hi). Here, we find clear partitioning between the different dynamic classes, with fast data exhibiting minimal $\sigma_\theta$, indicative of directed motion, and a large spread in $S_1$, indicative of wide-ranging structures; while the other classes generally exhibit higher $\sigma_\theta$ values correlated with lower and less varied $S_1$ values. These trends are consistent with the flow fields (Fig. 3Hii) and images (Fig. 3Hi inset).

Using BARCODE's IB branch we compare the initial maximum and secondary maximum island areas, $I_{0,1}$ and $I_{0,2}$, observing that slow and multi-mode composites exhibit predominantly a single large island, with $I_{0,2} < 5\%$ for all videos and $I_{0,1}$ values up to ~75% (Video 1 in Fig. 2 is an example). Fast class composites have a comparatively broader spread in initial island sizes and generally smaller $I_{0,1}$ and larger $I_{0,2}$ values. This trend is evident in the images (Fig. 3I inset) showing two large disconnected islands (circled data point $(I_{0,1}, I_{0,2}) \approx (37\%, 14\%)$), compared to Video 1 in Fig. 2 (boxed-in data point $(I_{0,1}, I_{0,2}) \approx (61\%, 0\%)$) that shows a fully connected network. Microtubules generally have smaller island sizes compared to actin, which may reflect increased network heterogeneity and larger mesh sizes.

If larger initial maximum island areas correlate with higher connectivity, then we would expect to see increased connectivity for slow and multi-mode compared to fast class materials, and for actin as compared to microtubules. We would also expect larger maximum void sizes to anticorrelate with connectivity, as a large enough void will inevitably break percolation. This is exactly observed in Fig. 3J, with nearly all slow and multi-mode networks remaining fully connected and most void areas below 50% of the FOV. Conversely, fast class data spans a wide range of connectivity values. Finally, the $C < 1$ data are predominantly for microtubules, which also have larger void areas ($V > 50\%$) compared to actin.

**BARCODE reveals universal features of active cytoskeleton materials**

To further validate BARCODE and demonstrate its rapid and robust characterization capabilities, we analyze two additional datasets: a published study that examined the contractile behavior of myosin-driven crosslinked actin [10] (Fig. 4A), and unpublished data on kinesin-driven composites of actin and microtubules, with and without crosslinking (Fig. 4H). In the first, confocal microscopy videos with separate channels representing labelled myosin and actin were collected. Previously published image analysis performed on the myosin channels revealed three classes of contractile behavior: local, global, and critically connected (Fig. 4A). Local contraction was signified by actin and myosin forming small scale (~10 µm) uniformly distributed clusters, whereas global contraction led all components to condense into a single well-defined region. Critically connected networks exhibited distinct reconfiguration across lengthscales, with some regions condensing to large aggregates while others remained more homogeneously distributed (Fig. 4A).

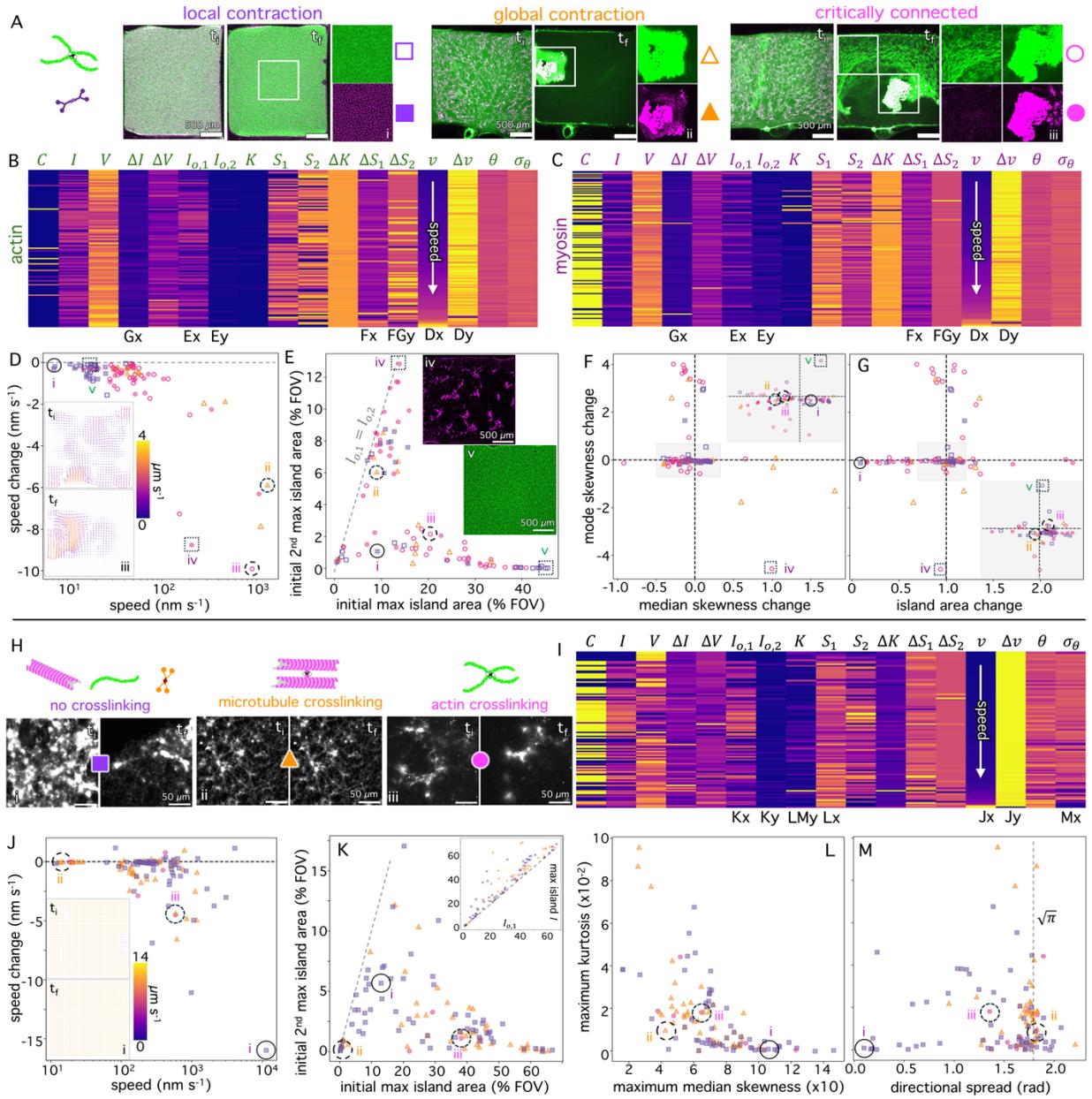

**Figure 4. BARCODE reveals universal dynamical and structural properties in diverse cytoskeletal networks.**
**(A-G)** Active networks of crosslinked actin (green schematic) and myosin (purple schematic), previously examined and classified using confocal fluorescence microscopy and image analysis as locally (purple) or globally (orange) contractile or critically connected (magenta)[10]. **(A)** Initial ($t_i$) and final ($t_f$) frames of representative multi-channel videos of crosslinked actin (green) and myosin (magenta) with single-channel zoom-ins of the boxed-in regions in $t_f$ frames. The colors and symbols shown for each class (local contraction: purple squares, global contraction: orange triangles, critically connected: magenta circles) and network component (actin: open, myosin: filled) are used in plots D-G. **(B,C)** Barcode matrix for the actin **(B)** and myosin **(C)** channels of the source videos [10], ordered by increasing mean speed. Letters below select columns denote the figure subpanel **(D-G)** and axis (x,y) where the metric is plotted. **(D)** Correlations between mean speed and speed change show similar dynamics for actin and myosin with locally contracting networks exhibiting the slowest speeds and smallest changes. Circled data i-iii correspond to the matching labeled video channels in (A). The first frames of the videos for boxed-in data points (iv,v) are shown in E. Inset: Downsampled (32×32 pixel) flow field for the initial and final frame of video channel iii in A (see SI Fig. S3 for full resolution RDS). **(E)** Correlation between initial maximum island area $I_{0,1}$ and initial secondary maximum island area $I_{0,2}$ shows distinct structural properties for actin (open) and myosin (filled) channels. Dashed line indicates $I_{0,1} = I_{0,2}$, above which no data can reside by definition. Myosin data near the line indicate more homogenous distribution of island sizes or clusters while most actin data fall along the horizontal near $I_{0,2} = 0$ suggesting one dominating contractile island. Inset: first frames of the myosin (iv) and actin (v) channels corresponding to matching labeled boxed-in data. **(F,G)** Correlations between mode skewness change and **(F)** median skewness change and **(G)** maximum island area change, with dashed lines denoting zero change. Grey inset plots are zoom-ins of the grey central regions that contain most of the data. The encircled and boxed-in data are the same as in D and E. Intensity distributions for circled and boxed data are shown in SI Fig. S4. Myosin and local contraction data display minimal restructuring (small change values) compared to actin and critical and global contraction classes. **(H-M)** BARCODE analysis of fluorescence confocal microscopy videos of networks comprising actin (green schematic), fluorescent-labeled microtubules (magenta schematic) and kinesin motors (orange schematic) with and without crosslinking of actin or microtubules via biotin-NA-biotin examined (see Methods). **(H)** Schematics of components and initial and final frames of example videos of labeled microtubules for the three compositions examined (no crosslinking: purple squares, microtubule crosslinking: orange triangles, actin crosslinking: magenta circles). **(I)** Barcode matrix for microtubule channel for all compositions ordered by increasing mean speed. Letters below select columns denote the figure subpanel and axis (x,y) where the metric is plotted. **(J)** Correlations between mean speed and speed change show similar trends as actomyosin networks in D. In all cases, contractile dynamics exhibit minimal or negative speed change, with uncrosslinked composites exhibiting the fastest speeds and largest decreases. Inset shows downsampled (32×32 pixel) flow fields for the initial and final frames of video i in H (see SI Fig. S5 for full resolution RDS). **(K)** Correlation between initial maximum island area $I_{0,1}$ and initial secondary maximum island area $I_{0,2}$ with dashed line indicating $I_{0,1} = I_{0,2}$. Inset shows $I_{0,1}$ versus maximum island area with the dashed line indicating their equality, suggesting that the largest island is primarily found at the beginning of each video, consistent with ongoing contraction. **(L,M)** Correlations between maximum kurtosis and **(L)** maximum median skewness and **(M)** flow directional spread, with dashed line in M denoting the expected spread of $\sigma_\theta = \sqrt{\pi}$ for randomly oriented motion. Circled data i-iii in plots J-M correspond to the matching labels in H. Intensity distributions for circled and boxed data are shown in SI Fig. S4.

To corroborate these general features of the different classes and discover their dynamical properties and structure-dynamics correlations, we use BARCODE to analyze both actin and myosin channels of the videos (Fig. 4B,C). Upon visual inspection of the barcodes, we find highly similar OF metrics between actin and myosin, suggesting that the two components are highly interacting and moving together. Both channels show larger maximum areas and changes for voids as compared to islands, while the actin channel shows a lower degree of connectivity and increased variability in skewness, suggesting more complex structure and reconfiguration, consistent with the images in Fig. 4A. We next analyze the quantitative metrics output by BARCODE, color-coding the data by the previously identified contractile classes (Fig. 4D-G).

Comparing the mean speed to the speed change (Fig. 4D), we confirm that myosin and actin dynamics are highly correlated. Locally contractile networks display low speeds on the order of ~10 nm s-1 with minimal changes, consistent with small-scale local contractions. By contrast, both critically connected and globally contractile networks exhibit higher and more variable speeds ranging from ~20 nm s-1 to ~103 nm s-1, with $v \gtrsim 10^2$ nm s-1 data showing increased slowing of dynamics. Moreover, BARCODE outputs indicate a coupling between dynamics and structure, as seen in comparing the RDS flow fields for the data with the highest speed change (Fig. 4D inset) with changing structures shown in Fig. 4Aiii. Higher speeds and their changes appear to be enabled by and result in more disconnected and heterogeneous structures (Fig. 4E inset).

To determine how the initial structure correlates with contractile class, we compare the initial maximum and secondary maximum island areas for actin and myosin (Fig. 4E). For actin, we find that locally contractile networks exhibit large $I_{0,1}$ values (>40%) and nearly zero $I_{0,2}$ values, similar to the slow class networks in Fig. 3I, and consistent with prior reports[10,20]. Conversely, the globally contractile and critically connected networks have smaller $I_{0,1}$ and larger $I_{0,2}$ values, similar to the fast networks in Fig. 3I. In contrast, the myosin data exhibits similar and generally correlated values of $I_{0,1}$ and $I_{0,2}$, with the critically connected networks having the highest ($I_{0,1}$, $I_{0,2}$) pairs, consistent with their expected clustering across different scales.

Finally, we relate restructuring to dynamics (Fig. 4F,G), comparing changes in mode skewness $\Delta S_2$ to changes in median skewness $\Delta S_1$ (Fig. 4F) and island area (Fig. 4G). We observe clear distinctions between the different contractile classes and between actin and myosin. We find that actin exhibits larger skewness changes compared to myosin, as visually indicated by the barcode, and global and critical contractile networks exhibit larger skewness changes compared to local contraction data, consistent with their expected contractile behavior. Interestingly, critical and global data have mostly negative and positive median skewness changes, respectively, indicating large-scale aggregation (brighter pixels) coupled with more network-poor (dark) regions for global data as compared to varied or multiple aggregates (brighter pixels) coupled with sustained network-rich regions (fewer dark pixels) for critically connected networks.

Another nontrivial relationship is that of the largest island area changes correlating with minimal mode skewness changes (Fig. 4G), indicating the image features are not changing substantially, maintaining the same intensity distribution, but rather they are reorganizing to increase or decrease island size. This result is consistent with the observation that the largest island area changes are observed in the myosin channel which overall have smaller island sizes and more clustering (Fig. 4E) resulting in amplified relative change in their size. These results highlight the power of BARCODE to reveal complex behaviors and correlations in active systems from a remarkably thrifty set of metrics produced in minutes.

Thus far, we have presented analysis of previously published and well-vetted data to benchmark BARCODE, revealing new insights and emergent similarities between two different cytoskeletal material. To explore potential universality in response, and further demonstrate BARCODE capabilities, we analyze unpublished data on kinesin-driven composites of actin and microtubules, with and without crosslinking (Fig. 4H). The system components are similar to those shown in Fig. 3A, but with a lower ratio of microtubules to actin, higher kinesin concentration, and no myosin (see Methods). The actin is also not labeled in this set of 109 videos, so our analysis focuses solely the structure and dynamics of the microtubule network. We observe a marked effect of crosslinking on dynamics and structure, with no crosslinking producing disparate clusters of microtubules that move in and out of the FOV, while crosslinked networks appear to be more connected and undergo more local contraction rather than large scale restructuring on the scale of the FOV (Fig. 4H). Comparing the barcodes for this system (Fig. 4I) to those of the actomyosin system (Fig. 4B), we find similar OF parameters, except $\Delta v$ appears higher for the kinesin-driven composites, and we observe much more variability in IB metrics. To further investigate similarities and differences between the different cytoskeleton systems, we evaluate similar correlations as in Figs 3 and 4D-G, color-coding the data by the type of crosslinking: none, microtubule-microtubule, or actin-actin.

We again find that slower speeds correlate with minimal speed changes and faster dynamics become increasingly slower over the course of a video ($\Delta v < 0$) (Fig. 4J). However, in comparison to actomyosin network dynamics, the speeds are generally an order of magnitude higher. Crosslinked microtubule networks exhibit the slowest speeds, with a cluster near $v \approx 20$ nm s-1, while most uncrosslinked networks display speeds of ~102 nm s-1 to >103 nm s-1, likely because crosslinking provides more rigidity and connectivity thereby slowing contraction. Increased connectivity should lead to larger differences between the initial maximum and secondary maximum island areas, since connected networks should have FOV-spanning islands, which we observe (Fig. 4K). Most of the crosslinked data lies below the equivalency line, indicating $I_{0,1} > I_{0,2}$, with a roughly inverse relationship between $I_{0,1}$ and $I_{0,2}$, similar to trends in Fig. 3I and the actin data in Fig. 4E. By contrast, the uncrosslinked network data falls near the equivalency line, indicating an ensemble of similarly sized clusters. Contraction is further confirmed by comparing $I_{0,1}$ with the maximum island area $I$ (Fig. 4K inset), which shows that most paired $I_{0,1}$ and $I$ values are similar, suggesting that islands are generally shrinking (i.e., contracting) over time.

To provide further insight into the network restructuring, we evaluate correlations between the maximum kurtosis $K$ and the maximum median skewness $S_1$ (Fig. 4L) and flow directional spread $\sigma_\theta$ (Fig. 4M). We find that $K$ and $S_1$ are generally inversely related and that most crosslinked networks exhibit higher kurtosis and lower skewness than uncrosslinked networks, which we attribute to the more heterogeneous nature of crosslinked networks that results in larger high-intensity tails and sharper low-intensity peaks. By contrast, uncrosslinked networks have more homogenously distributed bright clusters and single filaments that fill the FOV and sample a broader range of pixel intensities. We observe lower $\sigma_\theta$ in uncrosslinked networks (Fig. 4M), indicating more directed motion at relatively faster speeds (Fig. 4J), likely enabled by the reduced connectivity and rigidity compared to crosslinked networks. Conversely, most crosslinked network data exhibits directional spread of $\sigma_\theta \approx \sqrt{\pi}$, indicating randomly oriented motion. Together, these results demonstrate BARCODE's ability to quantify and correlate a host of structural and dynamic parameters for different active cytoskeleton network formulations, conditions, acquisition parameters, and spatiotemporal scales to reveal universal behaviors and emergent properties in this foundational class of active matter.

**BARCODE characterizes dynamics and structure of dynamic cell monolayers**

To further demonstrate the broad applicability of BARCODE, we next analyze two datasets of active cell monolayers. In each, large FOV (>1 mm) time-lapse videos are captured over the course of days (69-89 hrs) and the nuclei and cytoplasm are imaged separately, providing two channels for analysis. We first evaluate a previously published dataset [34] of videos of weakly interacting spindle-shaped human dermal fibroblasts (hdFs) (Fig. 5A) at two cell densities. The barcodes for each channel of the 72 videos (Fig. 5B) show both expected and nontrivial properties. For example, the entire set of nucleus data, which primarily contains isolated punctate objects that are not expected to significantly change shape or size, has essentially no connectivity, very small maximum and initial island sizes, and large voids. Conversely, the cytoplasm channel has more instances of positive connectivity along with larger maximum island areas and smaller voids. Despite clear structural differences, the dynamic OF metrics ($v, \Delta v, \theta, \sigma_\theta$) display similar trends between the channels, as expected since the two components comprise the same ensemble of cells. To more closely examine the effects of cell density, we evaluate similar correlations as in Figs 3 and 4.

Evaluating the speed and its change for each channel and cell density (Fig. 5C), we observe modest slowing for nearly all data, perhaps indicative of jamming as the cell number increases over time due to cell division, in line with the observation that the higher cell density data exhibits generally slower speeds. The cytoplasmic signal is generally slower than the nuclear signal, which we attribute to the more complex shape and intensity changes of the cytoplasm. Comparing the initial maximum and secondary maximum island areas, we find most of the data falling near the $I_{0,1} \approx I_{0,2}$ equivalency line. Moreover, the nucleus data is all tightly clustered around $I_{0,1} \approx I_{0,2} \approx 1\%$ which is equivalent to ~2000 µm2, roughly equivalent to the 2D projected area of the nucleus (Fig. 5A). The cytoplasm data displays more variation in island area and greater deviation from the equivalency line (i.e., $I_{0,1} > I_{0,2}$), indicating closely packed cells that register as larger multi-cell islands.

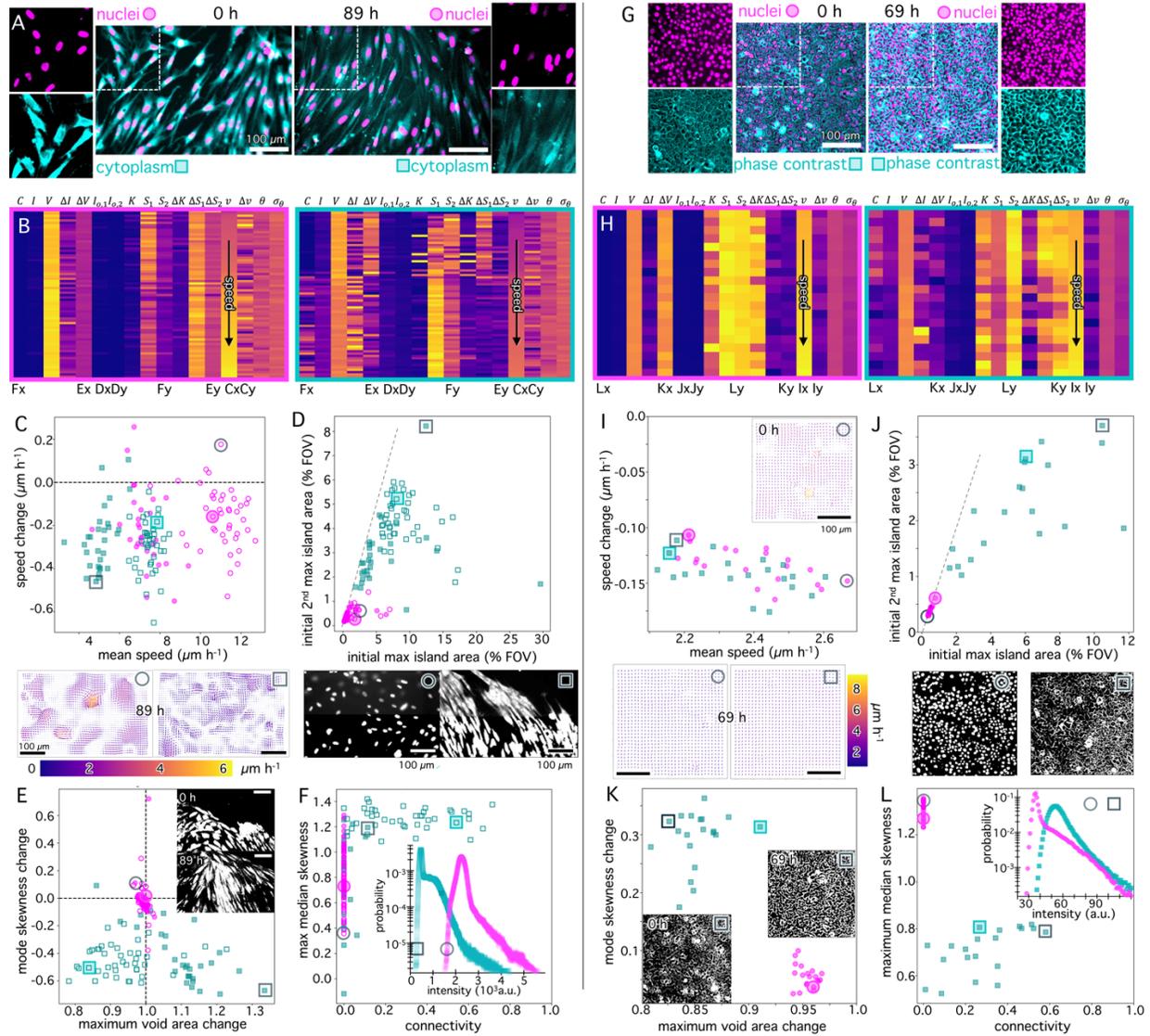

**Figure 5. BARCODE provides high-throughput characterization of dynamic cell monolayers. (A-F)** BARCODE analysis of epifluorescence microscopy videos of monolayers of human dermal fibroblast (hdF) cells with spectrally-distinct labeled nuclei (magenta, circles) and cytoplasm (cyan, squares) and two different cell concentrations [300 mm-2 (open), 450 mm-2 (filled)] as previously reported [34]. Scale bar is 100 µm for all images and RDS. **(A)** Initial (0 min) and final (89 min) frames of representative multi-channel video of low-density monolayers with single-channel zoom-ins of the boxed-in regions in each frame. **(B)** Barcode matrix for the nuclei (left, magenta border) and cytoplasm (right, cyan border) channels of the source videos [34] ordered by increasing mean speed. Letters below select columns denote the figure subpanel (C-F) and axis (x,y) where the metric is plotted. **(C-F)** Correlation plots with grey boxed-in and circled data corresponding to the cytoplasm and nuclei RDS shown below or as inset. Color-coded translucent filled boxes and circles enclose data points that correspond to the video shown in A. **(C)** (Top) Correlation between mean speed and speed change shows general slowing of dynamics and faster motion for lower concentration

monolayers. (Bottom) Final down-sampled (32×32 pixel) flow field RDS for circled and boxed data (see SI Fig. S6 for 8×8 RDS). **(D)** (Top) Correlation between initial maximum $I_{0,1}$ and secondary maximum $I_{0,2}$ island areas with dashed line indicating $I_{0,1} = I_{0,2}$. (Bottom) Binarization RDS (bottom half) and raw images (top half) for circled and boxed data. **(E)** Mode skewness change versus void area change with dashed lines denoting no change, and initial and final binarized images for boxed-in data shown. Initial and final intensity distributions for circled and boxed data are shown in SI Fig. S7. **(F)** Maximum median skewness $S_1$ versus connectivity $C$ with intensity distribution RDS shown for the frames at which $S_1$ is reached for the noted data. Intensity distributions for circled and boxed data are shown in SI Fig. S7. **(G-L)** BARCODE analysis of epifluorescence and phase contrast microscopy videos of monolayers of jamming MCF10A breast epithelial cells prepared as described [37,38]. Color and symbol keys, correlation plots and RDS mirror those in A-F. **(G)** Initial (0 min) and final (69 min) frames of representative multi-channel video of monolayers with single-channel zoom-ins of the boxed-in regions in each frame. **(H)** Barcode matrix for the nuclei and phase contrast channels, ordered by increasing mean speed. **(I)** Correlation between mean speed and speed change shows general slowing of dynamics and similar speeds for nuclei and phase contrast channels. Initial and final flow field RDS of circled data and final flow field for circled data, with 16×16 pixel down-sampling (see SI Fig. S8 for full resolution RDS). **(J)** Correlation between $I_{0,1}$ and $I_{0,2}$ with dashed line indicating $I_{0,1} = I_{0,2}$. Binarization RDS corresponding to noted conditions are shown. **(K)** Mode skewness change versus maximum void area change shows voids universally shrink and skewness increases, an effect that is amplified for phase contrast data and seen in inset binarization RDS. Initial and final intensity distributions for circled and boxed data are shown in SI Fig. S7. **(L)** Maximum median skewness $S_1$ versus connectivity $C$ with intensity distribution RDS shown in inset for the frames at which $S_1$ is reached for the noted data.

Turning to structural changes, quantified by $\Delta S_2$ and $\Delta V$ (Fig. 5E), we observe minimal changes in the nucleus channel, as expected, since they are not changing in size, shape or concentration. As expected, the cytoplasm data shows more variations in both metrics, reflecting changing cell shapes and clustering over time. Moreover, the mode skewness change is universally negative, indicative of increased uniformity and reduced spread in pixel intensities, which can be observed in Fig. 5A and S7. Interestingly, we find that the maximum void areas decrease for low density cases and increase for high densities. This clear density dependence suggests the presence of clustering without forming multi-cell islands for low density data, thereby reducing void areas, while cell clustering at higher densities result in large connected regions and similarly larger void areas. This physical picture is corroborated by Fig. 5F which shows that lower cell densities result in higher connectivity and generally higher median skewness, indicating a more connected meshwork of cells at low densities compared to large multi-cell aggregates that come with larger voids and more dark pixels.

To assess the relevance of these findings to other cell systems, we also analyze videos of jamming MCF10A breast epithelial cells (Fig. 5G-L). Here, whole cells and their GFP-labeled nuclei are imaged using phase contrast and epifluorescence, respectively (Fig. 5G). Examining the barcode for each channel (Fig. 5H), we find similarities with the hdFs (Fig. 5B), including minimal connectivity and island areas for the nucleus data and similar OF metrics between the two channels. Evaluating the same correlations as in Fig. 5C-F, we find the both cell types exhibit generally decreasing speeds (i.e., $\Delta v < 0$) (Fig. 5C,I) and similar $I_{0,1}$ vs $I_{0,2}$ correlations (Fig. 5D,J). However, the spread in $\Delta v$ and $v$ values is less for the MCF10A (Fig. 5I) compared to hdFs (Fig. 5C), suggesting more ordered and correlated motion within epithelia as compared to more motile fibroblasts. Moreover, the mode skewness change is positive ($\Delta S_2 > 0$) for epithelial cells (Fig. 5K) in contrast to hdFs and consistent with increased contrast and sharpness of final versus initial images (Fig. 5G,K), opposite to hdFs. As expected, and consistent with hdF data, we observe zero connectivity (Fig. 5F,L) and minimal void area change (Fig. 5E,K) for the nucleus channel and a wide range of connectivity values among the cell boundaries (Fig. 5F,L). Together, our results demonstrate the power of BARCODE to reveal convergent properties, dissect differences between disparate materials systems and accurately characterize material properties using distinct imaging modalities.

**DISCUSSION**

We present BARCODE, a HTP analysis platform that rapidly extracts a unique low-dimensional fingerprint from large and complex microscopy datasets of active materials. Each barcode output represents a readily accessible visual readout of 17 quantitative features, providing a concise and standardized summary of the system's structural and dynamic behavior and enabling rapid identification and comparison of properties among different experiments, formulations and conditions. In generating each barcode, the platform produces and archives standardized reduced data structures, including binarized frames, intensity distributions, and optical flow fields, to facilitate down-stream analysis, enabling users to perform deep-dives into datasets that exhibit irregularities or unexpected outputs. Our improved ability to analyze and share data is an important step towards democratizing materials discovery across disparate working groups with varied expertise.

We have demonstrated the effectiveness and efficiency of BARCODE in reproducing known material properties and discovering new features and formulation-structure-dynamics relationships. We have established BARCODE's abilities to analyze confocal fluorescence, epifluorescence and phase contrast videos ranging from 6 mins to 89 hours, 5 to 1178 frames, and with dimensions of 200 μm to 2 mm and 256 to 6318 pixels. The barcodes produced for each data set included multiple video channels, with individual file sizes as large as 6.7 GB and folder sizes as large as 205 GB. For all datasets, BARCODE analysis was completed in a range between 10 minutes for a 12 GB dataset to ~5 hours for a 205 GB dataset.

BARCODE is expressly designed to be a high throughput screening tool, rather than an intensive data analysis platform, to rapidly extract average and reduced-dimensionality metrics for coarse characterization to compare datasets, guide optimization and discover emergence. BARCODE requires minimal subjective inputs or training, and is executed with a user-friendly GUI, in contrast to algorithmically-intensive methods such as Fourier image analysis and particle-tracking methods, allowing for rapid analysis and broad use of the platform by beginners and experts alike [41,43-46]. In future applications, the rapid read-outs BARCODE provides may enable 'on-the-fly' training of machine learning tools for active material design and optimization, thereby unlocking the capabilities of complex multiphase multiscale soft materials in fit-for-purpose applications.

**ONLINE CONTENT**

Supplementary Information includes algorithmic BARCODE details (SI Section S1), extended data figures (SI Figures S1-S8) and tables (Tables S1, S2) referred to in the main text.

## METHODS

**Preparation and Imaging of Active Cytoskeleton Networks.** Data presented in Figures 2 and 3 were generated from previously published source videos [20]. Actomyosin data presented in Figure 4A-G were generated from previously published source videos [10]. The cytoskeletal networks and video data presented in Fig. 4H-M were generated as follows.

*Proteins:* All proteins were purchased as lyophilized powder from Cytoskeleton, Inc, reconstituted, flash-frozen into single-use aliquots, and stored at -80°C until use. Rabbit skeletal actin monomers (AKL99) and biotin-actin monomers (AB07) were reconstituted in G-buffer (2.0 mM Tris (pH 8), 0.2 mM ATP, 0.5 mM DTT, 0.1 mM CaCl2). Porcine brain tubulin dimers (T240), HiLyte 647-labeled tubulin dimers (TL670M) and biotin-tubulin dimers (T333P) were reconstituted in PEM-100 buffer (100 mM PIPES (pH 6.8), 2 mM MgCl2, and 2 mM EGTA).

Biotinylated kinesin-401 [47,48] was expressed in Rosetta (DE3) pLysS competent E. coli cells (ThermoFisher), purified, and flash-frozen into single-use aliquots, as described previously [20]. To prepare force-generating kinesin clusters, kinesin-401 dimers were incubated with NeutrAvidin (ThermoFisher) at a 1.7:1 ratio in PEM-100 to a final concentration of 10 µM supplemented with 4 mM DTT for 30 min at 4°C. Clusters were prepared fresh and used within 24 hrs.

For composites that incorporate actin or microtubule crosslinking, actin:actin or microtubule:microtubule crosslinked complexes were prepared according to previously described protocols [49]. In brief, biotinylated actin monomers or tubulin dimers were combined with NeutrAvidin and free biotin at a ratio of 2:2:1 protein:free biotin:NeutrAvidin.

*Cytoskeletal Network Preparation:* Actin-microtubule composites were prepared by polymerizing a mixture of 5.22 µM actin monomers, 6.06 µM tubulin dimers, 0.32 µM HiLyte 647-labeled tubulin dimers, 5.22 µM phalloidin (Invitrogen) and 5 µM Taxol (Sigma) in PEM-100, supplemented with 0.1% Tween, 4 mM ATP, and 4 mM GTP. For crosslinked composites, a portion of either actin monomers or tubulin dimers was replaced with equivalent crosslinker complexes to achieve the same overall actin and tubulin concentrations and crosslinker:protein ratios of $R_A = 0.02$ for actin or $R_T = 0.005$ for tubulin. Composites were polymerized in the dark for 1 hour at 37°C.

To prepare ACCs, 5 µL of the polymerized actin-microtubule composite was combined with the following to a 9 µL final volume: oxygen scavenging system [45 µg/mL glucose, 0.005% β-mercaptoethanol, 43 µg/mL glucose oxidase, 7 µg/mL catalase, 2 mM Trolox (Sigma)] and ATP-regeneration system [26.7 mM phosphoenol pyruvate (Beantown Chemical, 129745) and pyruvate kinase/lactate dehydrogenase (Sigma, P-029)]. Finally, 1 µL of kinesin clusters was added to reach a final concentration of 1 µM, followed by gentle mixing of the sample by pipetting up and down.

*Sample chambers*: Sample chambers with a volume of ~10 µL were made by placing two strips of parafilm between a No. 1 glass coverslip and microscope slide, followed by heating to fuse them together. To prevent surface adsorption of proteins, the chambers were filled with a 1 mM solution of BSA (bovine serum albumin, Sigma) and incubated for 10 minutes, after which the solution is flushed out with compressed air. The prepared cytoskeletal network solution was then loaded into the chamber and the open ends were sealed with UV-curable glue.

*Imaging*: Imaging of HiLyte647-labeled microtubules was performed using a Nikon A1R laser scanning confocal microscope with a 60× oil-immersion objective (Nikon) and a 640 nm laser with 624±20 nm / 692±20 nm excitation/emission filters. Time-series (videos) of 256 × 256 square-pixel (213 μm × 213 μm) images were collected at 1.33 fps for a minimum of 400 frames (300 s). Imaging began 5 minutes after the addition of kinesin motors to the sample in the middle plane of the ~100 μm thick sample chamber. Each subsequent video was recorded in a different field of view laterally translated by at least 500 μm. Imaging continued until restructuring or motion was no longer visible (~60-120 min). 5-10 videos were collected for each sample. Each data point shown in Fig. 4 corresponds to a single video. The 55, 7 and 47 data points for the uncrosslinked, actin crosslinked and microtubule crosslinked data, respectively are from 8, 2 and 6 independent replicates.

**Preparation and Imaging of Cell Monolayers.** Data presented in Figure 5A-F were generated from source videos [34] which we further processed as described below. Video data presented in Fig. 5G-J, from experiments similar to previous reports[37,38], were graciously shared by Jasmin Di Franco and Roberto Cerbino (University of Vienna, Austria) and prepared as described below.

*Cell culture:* MCF10A cells were cultured in Dulbecco's Modified Eagle Medium: Nutrient Mixture F-12 supplement with Glutamine (DMEM/F12 GlutaMax) medium (Gibco), supplemented with 5% horse serum (Biowest), 0.5 mg/mL hydrocortisone (Sigma-Aldrich), 100 ng/mL cholera toxin (Sigma-Aldrich), 10 μg/mL insulin (Roche), 1% Penicillin-Streptomycin (HyClone), and 20 ng/mL EGF (Peprotech), the latter being added directly to the culture plates. Cells were maintained at 37°C in a humidified atmosphere with 5% $CO_2$. Stable expression of GFP-H2B was achieved by lentiviral infection of MCF10A cells with pBABE-puro-GFP-H2B vectors to enable nuclear labeling.

*Cell jamming assay:* Cells were seeded into six-well plates at a density of $1.5 \times 10^6$ cells per well in complete medium and cultured to form a uniform monolayer (~24 hours). Prior to imaging, the cell monolayer was carefully washed with 1X Dulbecco's Phosphate Buffered Saline (DPBS) to remove floating cells, and the medium was refreshed. Time-lapse images of size 1024 ×1024 square-pixels (1331 μm × 1331 μm) were captured every 5 minutes over a 72-hour period using a Leica Thunder inverted microscope equipped with a 10× objective, both in phase contrast, to image the cell walls and cytoplasm, as well as fluorescence, to image the GFP-labeled nuclei. The assay was conducted in an environmental microscope incubator set to 37°C with 5% $CO_2$ perfusion. Data presented in Fig. 5G-J are from 20 videos from 5 independent samples, each containing a fluorescence and phase contrast channel. For BARCODE analysis, we divided each video channel into four 512 ×512 square-pixel images to increase statistics.

*Fig. 5A-F video post-processing:* Videos [34] of size 6318×3546 square-pixels (3702 μm × 2078 μm) and 5088 × 2332 square-pixels (2982 μm × 1367 μm) were divided into smaller FOVs to increase statistics and remove spurious dark spots or sample borders that impact image analysis. Specifically, the data is from 72 videos from 2 independent samples, each containing two channels. For BARCODE analysis, we divided each video channel into 30-42 video tiles with tile size of 960×608 square-pixels (562 μm × 356 μm).

## BARCODE Workflow

The algorithmic workflow for BARCODE is described in the main text and the Supplemental Information (SI, Section 1). BARCODE was developed using Python 3.12 and uses the packages nd2 and ImageIO for reading .tif and .nd2 files into an array of dimensions (T, X, Y, C), where T is the number of frames, X and Y representing the width and height of the video, and C is the number of channels. The packages Numpy, OpenCV, Scikit-Image, Scipy, and Matplotlib are used for data processing and visualization. To develop the graphical user interface, we used the Python package Gooey[50], with software packaging performed using the package PyInstaller. BARCODE processes data in the form of .tif and .nd2 files, and produces outputs in the form of RDS structures saved as .csv files, and a .csv barcode file with the metrics derived from the RDS structures, with an option to also output a colorized .svg barcode (e.g., Figs. 1E, 2E).

**BARCODE Parameters**

**Table 1.** Summary of BARCODE parameters, fully described in the main text and SI Section 1.

| Parameter | Description |
| --- | --- |
| Connectivity $C$ | Fraction of frames in a video in which material is percolated across at least one dimension |
| Maximum Island Area $I$ | Area (fraction of area of FOV) of largest contiguous region of white pixels across all frames of a binarized video |
| Maximum Void Area $V$ | Area (fraction of area of FOV) of largest contiguous region of black pixels across all frames of a binarized video |
| Island Area Change $\Delta I$ | Relative change in maximum island area over the time course of the video |
| Void Area Change $\Delta V$ | Relative change in maximum void area over the time course of the video |
| Initial Maximum Island Area $I_{0,1}$ | Area of largest island at start of video |
| Initial 2nd Maximum Island Area $I_{0,2}$ | Area of second largest island at start of video |
| Maximum Kurtosis $K$ | Maximum kurtosis value measured using the initial and final 10% of frames |
| Kurtosis Change $\Delta K$ | Change in maximum kurtosis value between final and initial 10% of frames in a video |
| Maximum Median Skewness $S_1$ | Maximum median skewness value measured using the first and last 10% of frames |
| Median Skewness Change $\Delta S_1$ | Change in maximum median skewness value between final and initial 10% of frames in a video |
| Maximum Mode Skewness $S_2$ | Maximum mode skewness value measured using the initial and final 10% of frames |
| Mode Skewness Change $\Delta S_2$ | Change in maximum mode skewness value between final and initial 10% of frames in a video |
| Mean Speed $v$ | Mean speed of material across all vectors of all flow fields |
| Speed Change $\Delta v$ | Difference between the mean speed measured in the initial and final flow field |
| Mean Flow Direction $\theta$ | Average direction of material motion over all flow fields |
| Directional Spread $\sigma_\theta$ | Standard deviation of motion direction over all flow fields |

## DATA AVAILABILITY

The source data for all plots shown in Figures 2-5 can be found on Zenodo at doi: 10.5281/zenodo.14641721

## CODE AVAILABILITY

BARCODE and supporting documentation are available on the GitHub repository at https://github.com/softmatterdb/barcode

## ACKNOWLEDGEMENTS

BARCODE was developed through Hackathon events held at University of California, Santa Barbara and supported by the US National Science Foundation (NSF) Designing Materials to Revolutionize and Engineer our Future (DMREF) program with contributions from the following participants: Jonathan Michel, Mengyang Gu, Prashali Chauhan, Laura Morocho, Nimisha Krishnan, Anindya Chowdhury, Lauren Melcher, JJ Siu, Gregor Leech, Mehrzad Sasanpour, and Karthik Peddireddy. We acknowledge funding from the US National Science Foundation DMREF program through following grants: NSF DMR-2119663 (to RMRA), NSF DMR-2118403 (to JLR), NSF DMR-2118449 (to MD), NSF DMR-2118497 (to MTV), Research Corporation for Science Advancement award no. CS-PBP-2023-019 (to RMRA, MH), Arnold and Mabel Beckman Foundation Beckman Scholars Program (to RMRA, KM) and the NSF BioPACIFIC Materials Innovation Platform NSF DMR-1933487 for personnel support (QC) and access to research infrastructure. We thank Christopher Dunham, BioPACIFIC MIP, Emmie Kao and Christopher Tao for helpful discussions and preliminary research development. We thank Jose Alvarado, Gijsje Koenderink, and Yimin Luo for providing data [10,34] and for helpful discussions. We thank Roberto Cerbino and Jasmin Di Franco for providing unpublished data and for helpful discussions.

## AUTHOR CONTRIBUTIONS

MTV and RMRA conceived and designed the research. QC, AS, AD, KM, MH, KT, RM, RMRA and MTV designed BARCODE algorithms, pipeline, and documentation. KM and MH performed experimental work and curated data. QC, AS, AD, RMRA and MTV analyzed data. QC, AS, RMRA, and MTV prepared figures and wrote the manuscript. MD, RMRA, and MTV supervised the research. All authors interpreted data and edited the manuscript.

**BARCODE: Biomaterial Activity Readouts to Categorize, Optimize, Design and Engineer for high throughput screening and characterization of dynamically restructuring soft materials**

Qiaopeng Chen†[1], Aditya Sriram†[2], Ayan Das[1], Katarina Matic[2], Maya Hendija[2], Keegan Tonry[3], Jennifer L. Ross[4], Moumita Das[3], Ryan J. McGorty[2], Rae M. Robertson-Anderson[2*], Megan T. Valentine[1*]

[1] Department of Mechanical Engineering, University of California, Santa Barbara
[2] Department of Physics and Biophysics, University of San Diego
[3] School of Physics and Astronomy, Rochester Institute of Technology
[4] Department of Physics, Syracuse University

\* Corresponding Authors, equal contributions
† Equal Contribution

## Supplemental Information

**Section S1. Detailed description and formulas for BARCODE parameters**

**Table S1.** BARCODE run time and efficiency metrics for the four different datasets examined in the main text.

**Table S2.** Statistical analysis of agreement between filaments speeds computed with BARCODE and those reported in doi:10.1093/pnasnexus/pgad245.

**Figure S1.** Intensity probability distributions for the circled data in Figure 3H in the main text.

**Figure S2.** Velocity vector flow field RDS for the circled data in Figure 3H in the main text.

**Figure S3.** Velocity vector flow field RDS for the circled iii data in Figure 4D in the main text.

**Figure S4.** Intensity probability distributions for the enumerated data points in Figure 4L (left) and 4F (right) in the main text.

**Figure S5.** Velocity vector flow field RDS for the circled i data in Figure 4J in the main text.

**Figure S6.** Velocity vector flow field RDS for the circled data in Figure 5C in the main text

**Figure S7.** Intensity probability distributions for the circled data points in Figure 5E (left) and 4F (right) in the main text.

**Figure S8.** Velocity vector flow field RDS for the circled data in Figure 5I in the main text.

**Table S1. BARCODE run time and efficiency metrics for the four different datasets examined in the main text.**

| Dataset | active cytoskeleton composite | actomyosin network | kinesin-driven composite | hdF cell monolayer | MCF10A cell monolayer |
|---|---|---|---|---|---|
| source | * | ** | see methods | *** | see methods |
| barcode | Fig 3C,D | Fig 4B,C | Fig 4I | Fig 5B | Fig 5H |
| # of videos | 48 | 132 | 131 | 72 | 20 |
| # of channels | 2 | 2 | 1 | 2 | 2 |
| avg. frames per video | 979 | 394 | 265 | 117 | 829 |
| frame size (pixels) | 256 x 256 | 1024 x 1024 | 256 x 256 | 960 x 608 | 512 x 512 |
| avg. video size (MB) | 123 | 775 | 364 | 131 | 207 |
| dataset size (GB) | 11.8 | 200 | 44.6 | 18.4 | 8.1 |
| barcode run time (mins) | 10 | 318 | 35 | 120 | 16 |
| avg. run time per video (s) | 5.6 | 72 | 19 | 50 | 25 |
| avg. run time per GB (s) | 46 | 114 | 70 | 377 | 120 |

*doi:10.1093/pnasnexus/pgad245

**doi:10.1038/nphys2715

***doi:10.1098/rsif.2023.0160

**Table S2. Statistical analysis of agreement between filaments speeds computed with BARCODE and those reported in doi:10.1093/pnasnexus/pgad245.** Analysis of the Pearson correlation coefficient shows a high degree of statistical agreement between the two datasets for all channels (actin, microtubule) and dynamical classes (slow, fast, multi-mode).

| data class | all | slow | fast | multi-mode |
|---|---|---|---|---|
| **both channels** | | | | |
| Pearson coefficient | 0.99 | 0.87 | 0.99 | 0.99 |
| p-value | < 0.01 | < 0.01 | < 0.01 | < 0.01 |
| **actin channel** | | | | |
| Pearson coefficient | 0.98 | 0.89 | 0.98 | 0.99 |
| p-value | < 0.01 | < 0.01 | < 0.01 | < 0.01 |
| **microtubule channel** | | | | |
| Pearson coefficient | 0.99 | 0.86 | 0.99 | 0.99 |
| p-value | < 0.01 | < 0.01 | < 0.01 | < 0.01 |

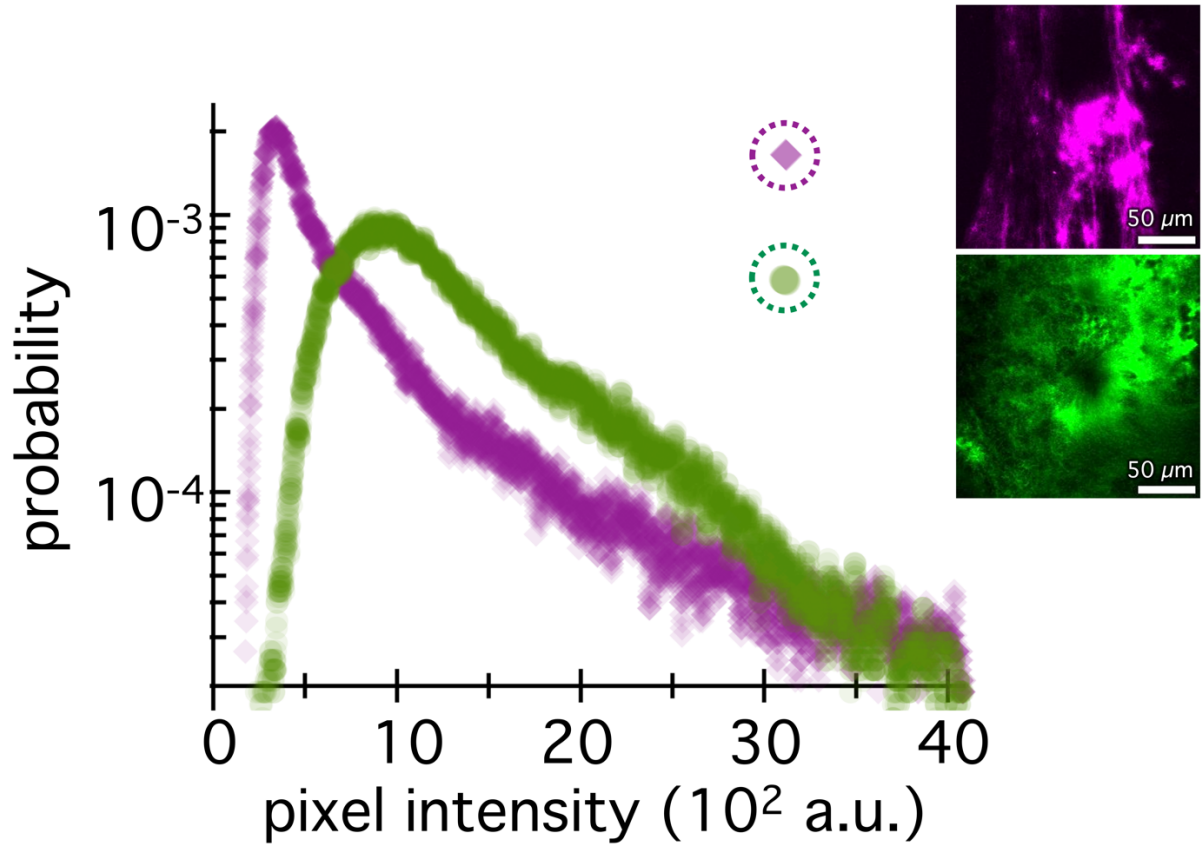

**Figure S1. Intensity probability distributions for the circled data in Figure 3H in the main text.** The distributions are for the frame at which the median skewness is maximum ($S_1$) for the data circled in green and magenta in the Figure 3H in the main text. The green and magenta images are the corresponding frames for the color-matched distributions.

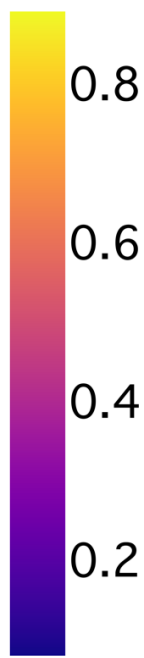
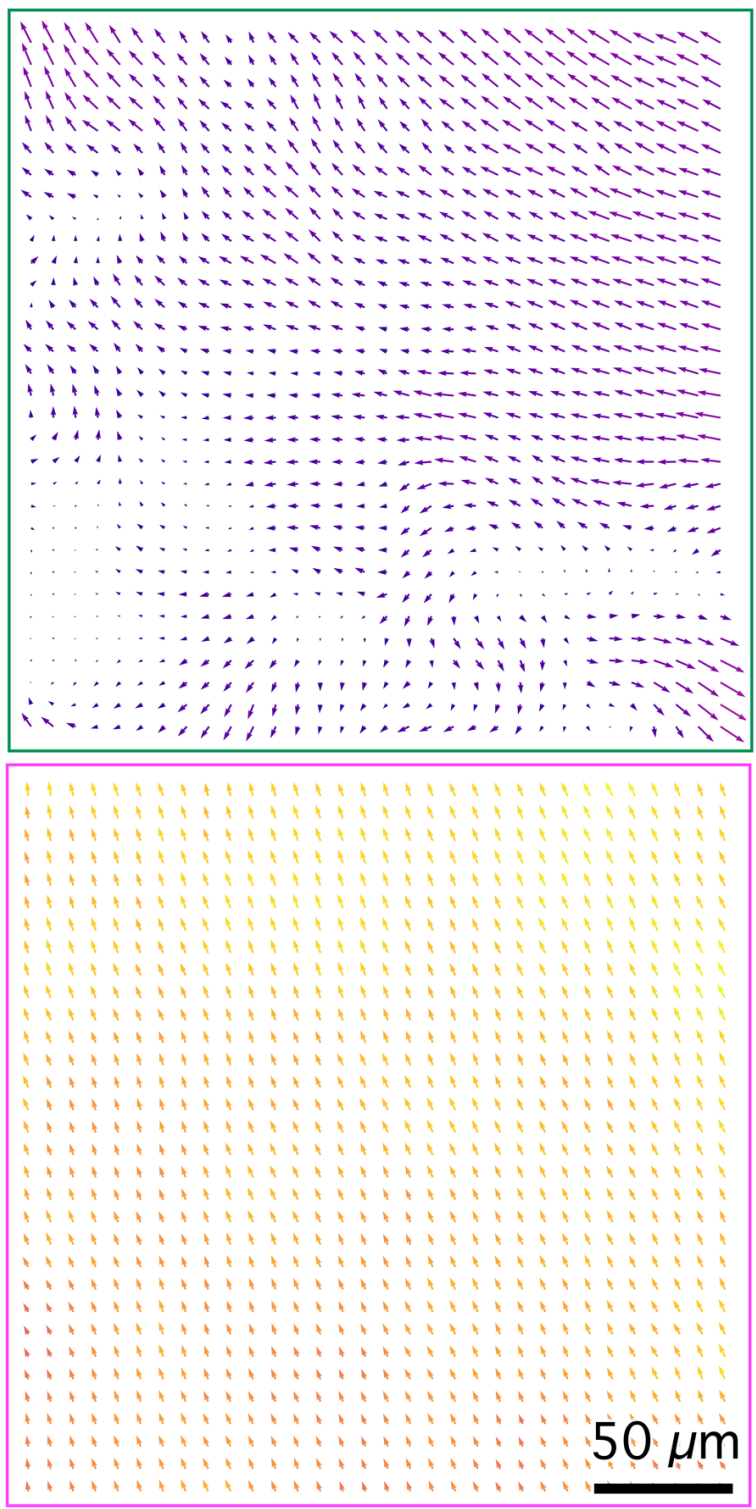

**Figure S2. Velocity vector flow field RDS for the circled data in Figure 3H in the main text.** The full resolution flow field RDS generated with BARCODE from which the down-sampled flow fields shown in Figure 3H.ii are generated. The arrow color and size indicate the speed. The arrow color is normalized relative to both flow fields, whereas the arrow size is scaled relative to the single field it represents. The direction in which the arrow points indicates the velocity direction at each point.

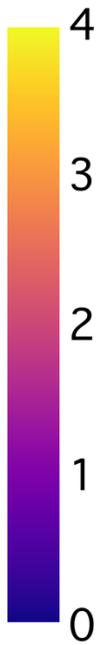
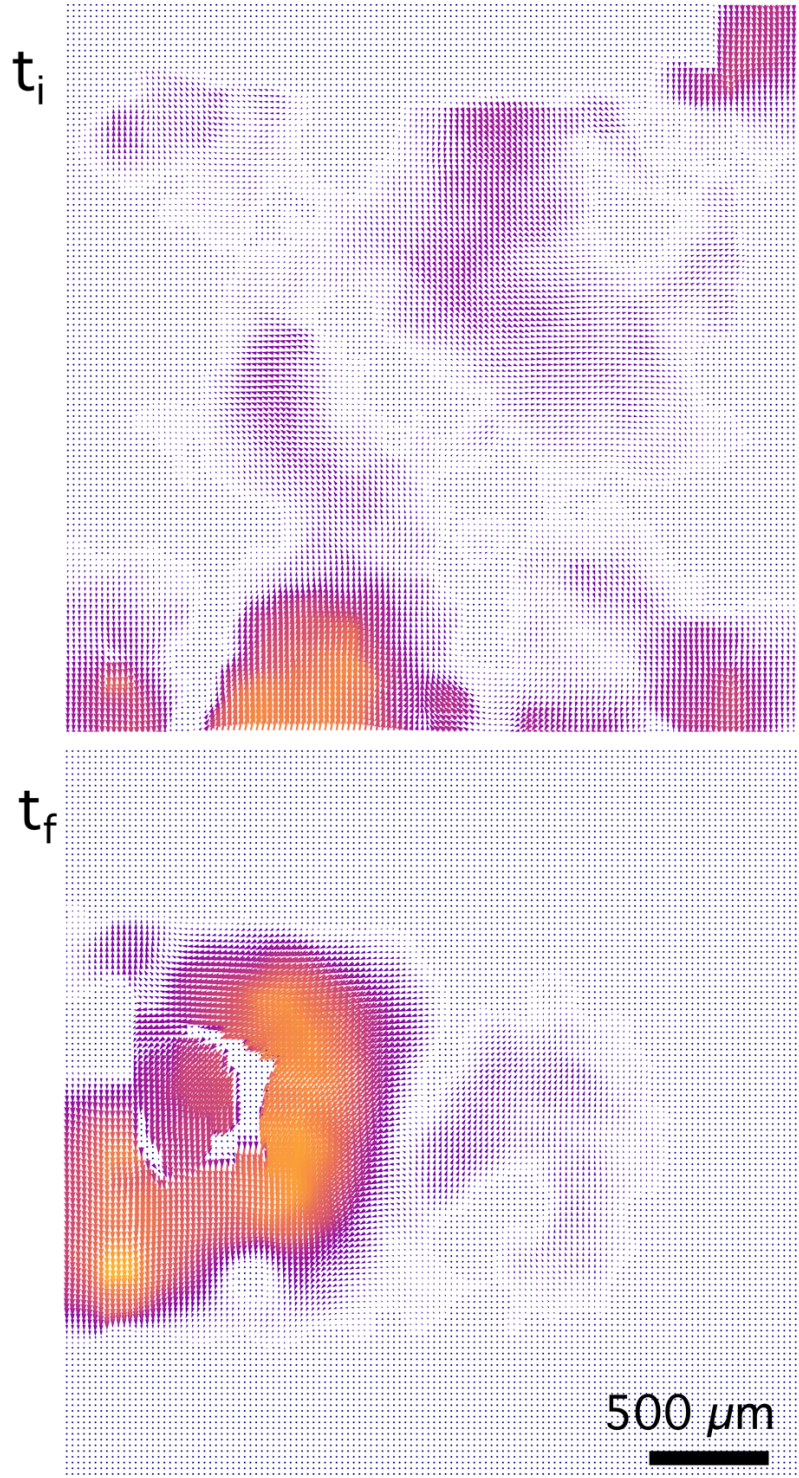

**Figure S3. Velocity vector flow field RDS for the circled iii data in Figure 4D in the main text.** The full resolution flow field RDS generated with BARCODE from which the down-sampled flow fields shown in Figure 4D are generated. The arrow color and size indicate the speed. The arrow color is normalized relative to both flow fields, whereas the arrow size is scaled relative to the single field it represents. The direction in which the arrow points indicates the velocity direction at each point. In regions of very high speed the arrows overlap to a saturating limit so the direction is not discernible at the full resolution.

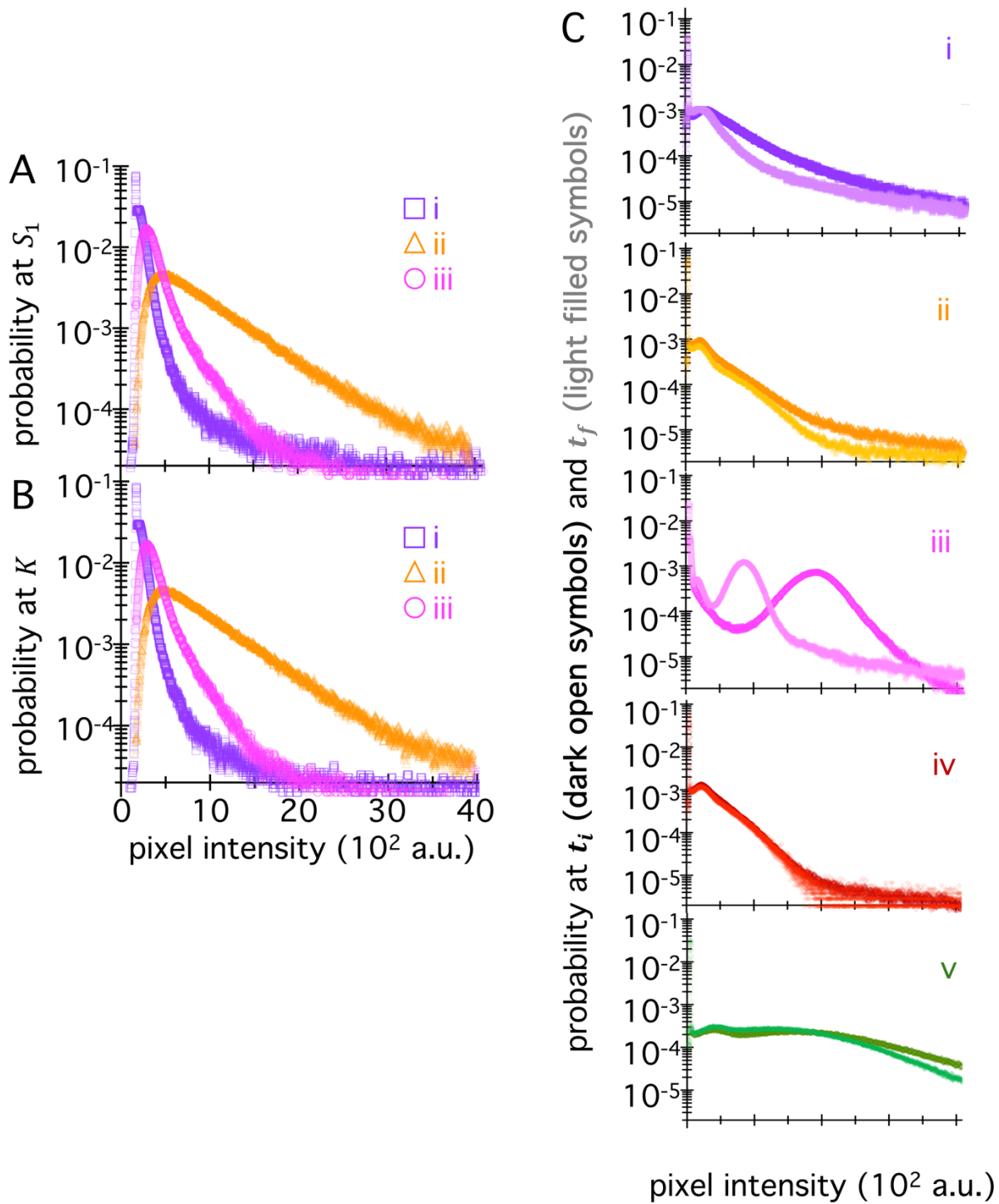

**Figure S4. Intensity probability distributions for the enumerated data points in Figure 4L (left) and 4F (right) in the main text. (A,B)** The distribution for the frame that has the (A) maximum median skewness and (B) maximum kurtosis for the circled data in Figure 4L in the main text labeled as i (purple squares), ii (orange triangles), iii (magenta circles). **(C)** Intensity probability distributions for the initial (dark open symbols) and final (light filled symbols) frames for the data points in Figure 4F that are circled (i-iii) or boxed-in (iv, v).

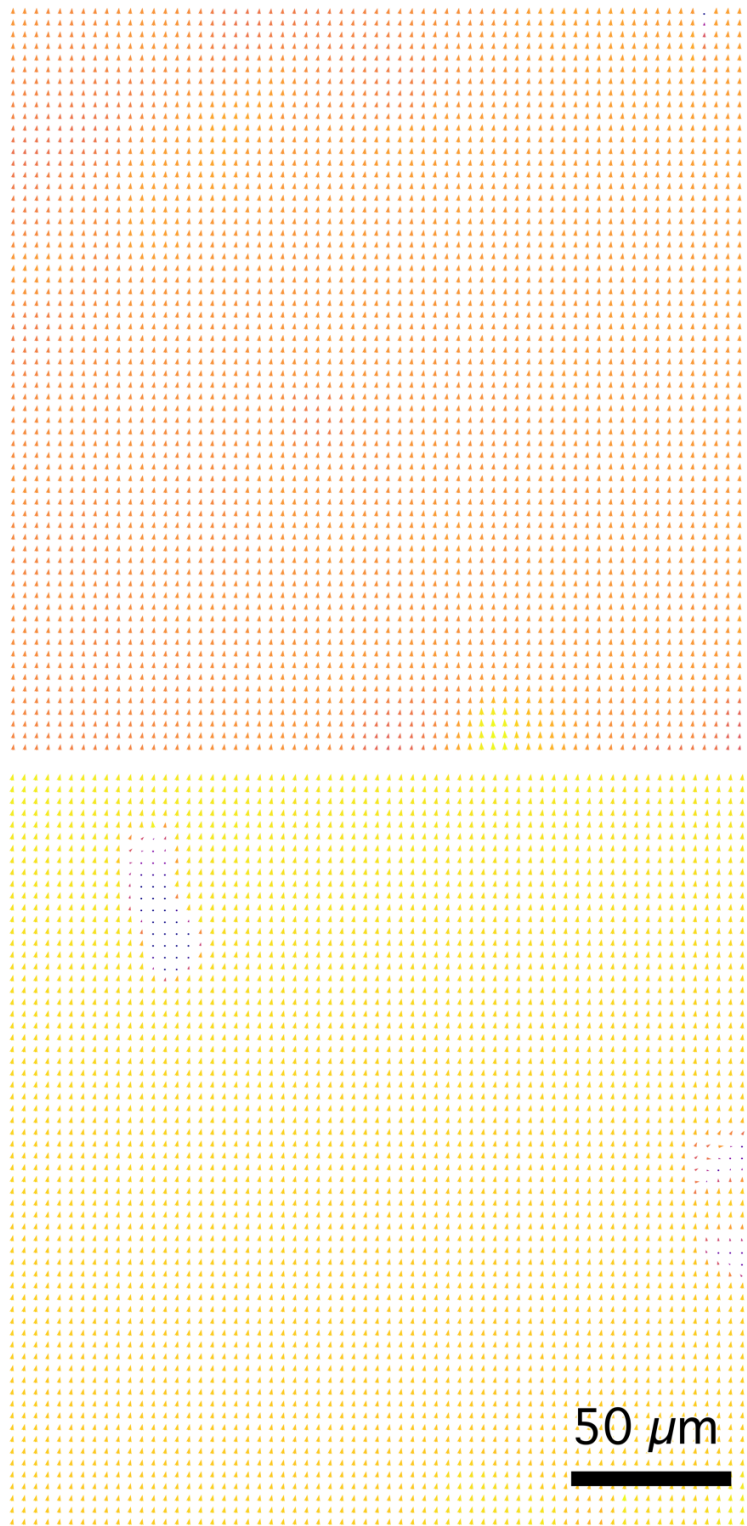
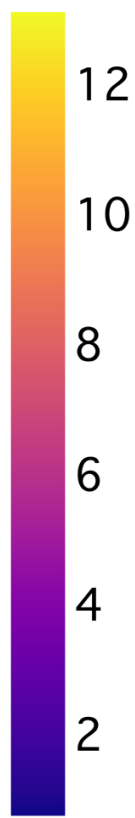

**Figure S5. Velocity vector flow field RDS for the circled i data in Figure 4J in the main text.** The full resolution flow field RDS generated with BARCODE from which the down-sampled flow fields shown in Figure 4J are generated. The arrow color and size indicate the speed. The arrow color is normalized relative to both flow fields, whereas the arrow size is scaled relative to the single field it represents. The direction in which the arrow points indicates the velocity direction at each point.

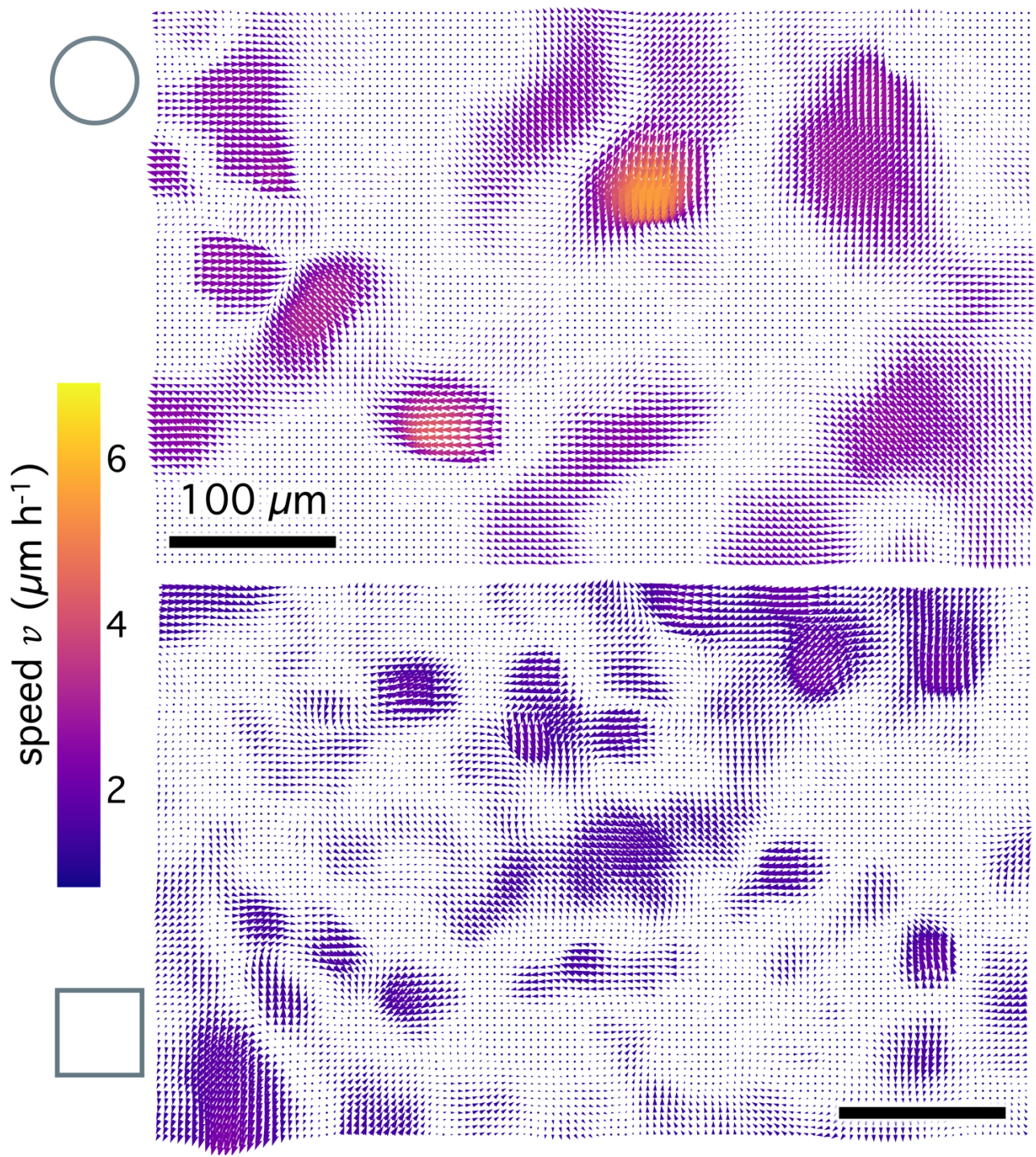

**Figure S6. Velocity vector flow field RDS for the circled data in Figure 5C in the main text.** The full resolution flow field RDS generated with BARCODE from which the down-sampled flow fields shown in Figure 5C are generated. The arrow color and size indicate the speed. The arrow color is normalized relative to both flow fields, whereas the arrow size is scaled relative to the single field it represents. The direction in which the arrow points indicates the velocity direction at each point.

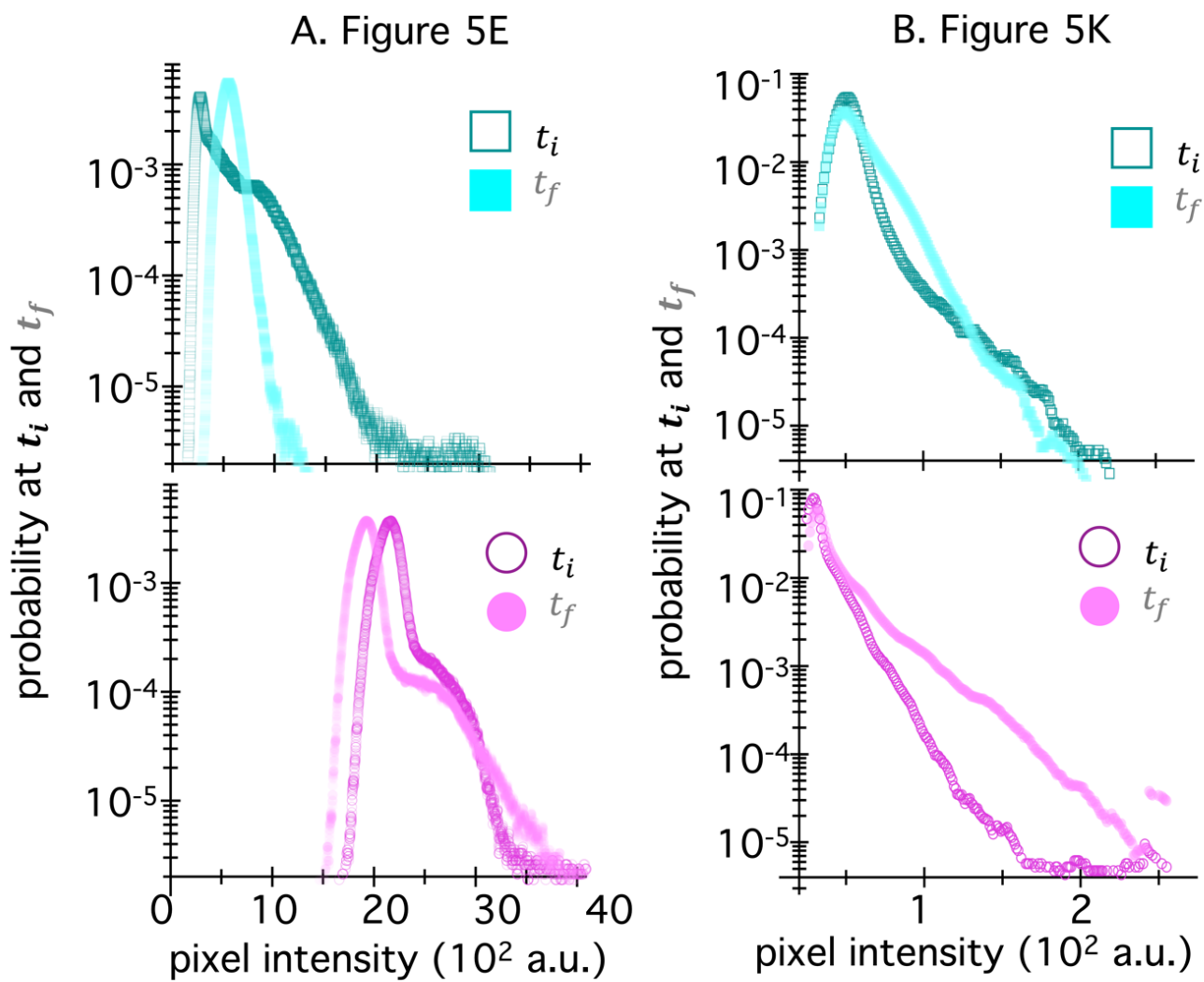

**Figure S7. Intensity probability distributions for the circled data points in Figure 5E (left) and 5K (right) in the main text.** Intensity probability distributions for the initial (dark open symbols) and final (light filled symbols) frames for the circled data points in (A) Figure 5E and (B) Figure 5K. Top plots correspond to the (A) cytoplasm or (B) phase contrast channel data points and bottom plots are for the circled nucleus channel data points.

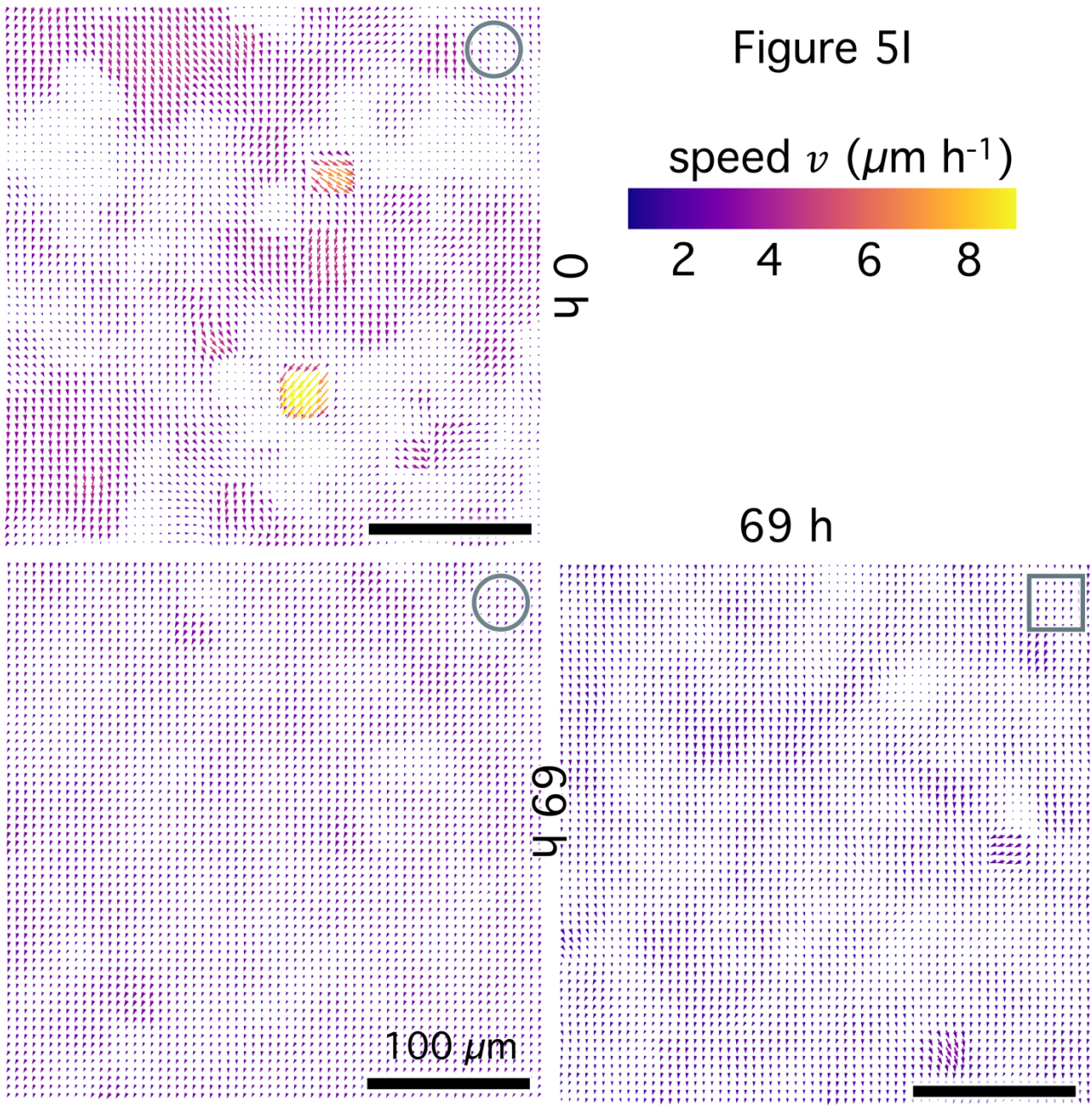

**Figure S8. Velocity vector flow field RDS for the circled data in Figure 5I in the main text.** The full resolution flow field RDS generated with BARCODE from which the down-sampled flow fields shown in Figure 5I are generated. The arrow color and size indicate the speed. The arrow color is normalized relative to both flow fields, whereas the arrow size is scaled relative to the single field it represents. The direction in which the arrow points indicates the velocity direction at each point.

**Table S1. BARCODE run time and efficiency metrics for the four different datasets examined in the main text.**

| Dataset | active cytoskeleton composite | actomyosin network | kinesin-driven composite | hdF cell monolayer | MCF10A cell monolayer |
|---|---|---|---|---|---|
| source | * | ** | see methods | *** | see methods |
| barcode | Fig 3C,D | Fig 4B,C | Fig 4I | Fig 5B | Fig 5H |
| # of videos | 48 | 132 | 131 | 72 | 20 |
| # of channels | 2 | 2 | 1 | 2 | 2 |
| avg. frames per video | 979 | 394 | 265 | 117 | 829 |
| frame size (pixels) | 256 x 256 | 1024 x 1024 | 256 x 256 | 960 x 608 | 512 x 512 |
| avg. video size (MB) | 123 | 775 | 364 | 131 | 207 |
| dataset size (GB) | 11.8 | 200 | 44.6 | 18.4 | 8.1 |
| barcode run time (mins) | 10 | 318 | 35 | 120 | 16 |
| avg. run time per video (s) | 5.6 | 72 | 19 | 50 | 25 |
| avg. run time per GB (s) | 46 | 114 | 70 | 377 | 120 |

*doi:10.1093/pnasnexus/pgad245

**doi:10.1038/nphys2715

***doi:10.1098/rsif.2023.0160

**Table S2. Statistical analysis of agreement between filaments speeds computed with BARCODE and those reported in doi:10.1093/pnasnexus/pgad245.** Analysis of the Pearson correlation coefficient shows a high degree of statistical agreement between the two datasets for all channels (actin, microtubule) and dynamical classes (slow, fast, multi-mode).

| data class | all | slow | fast | multi-mode |
|---|---|---|---|---|
| **both channels** | | | | |
| Pearson coefficient | 0.99 | 0.87 | 0.99 | 0.99 |
| p-value | < 0.01 | < 0.01 | < 0.01 | < 0.01 |
| **actin channel** | | | | |
| Pearson coefficient | 0.98 | 0.89 | 0.98 | 0.99 |
| p-value | < 0.01 | < 0.01 | < 0.01 | < 0.01 |
| **microtubule channel** | | | | |
| Pearson coefficient | 0.99 | 0.86 | 0.99 | 0.99 |
| p-value | < 0.01 | < 0.01 | < 0.01 | < 0.01 |

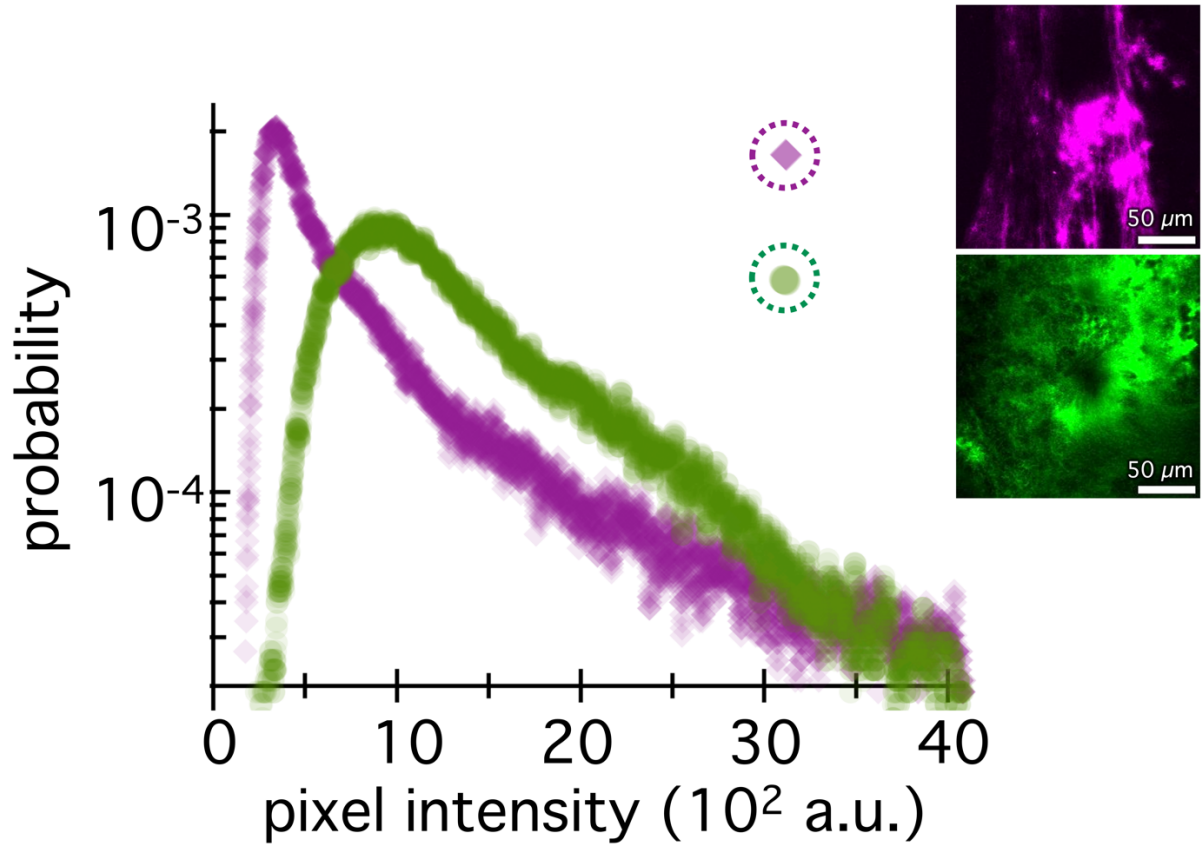

**Figure S1. Intensity probability distributions for the circled data in Figure 3H in the main text.** The distributions are for the frame at which the median skewness is maximum ($S_1$) for the data circled in green and magenta in the Figure 3H in the main text. The green and magenta images are the corresponding frames for the color-matched distributions.

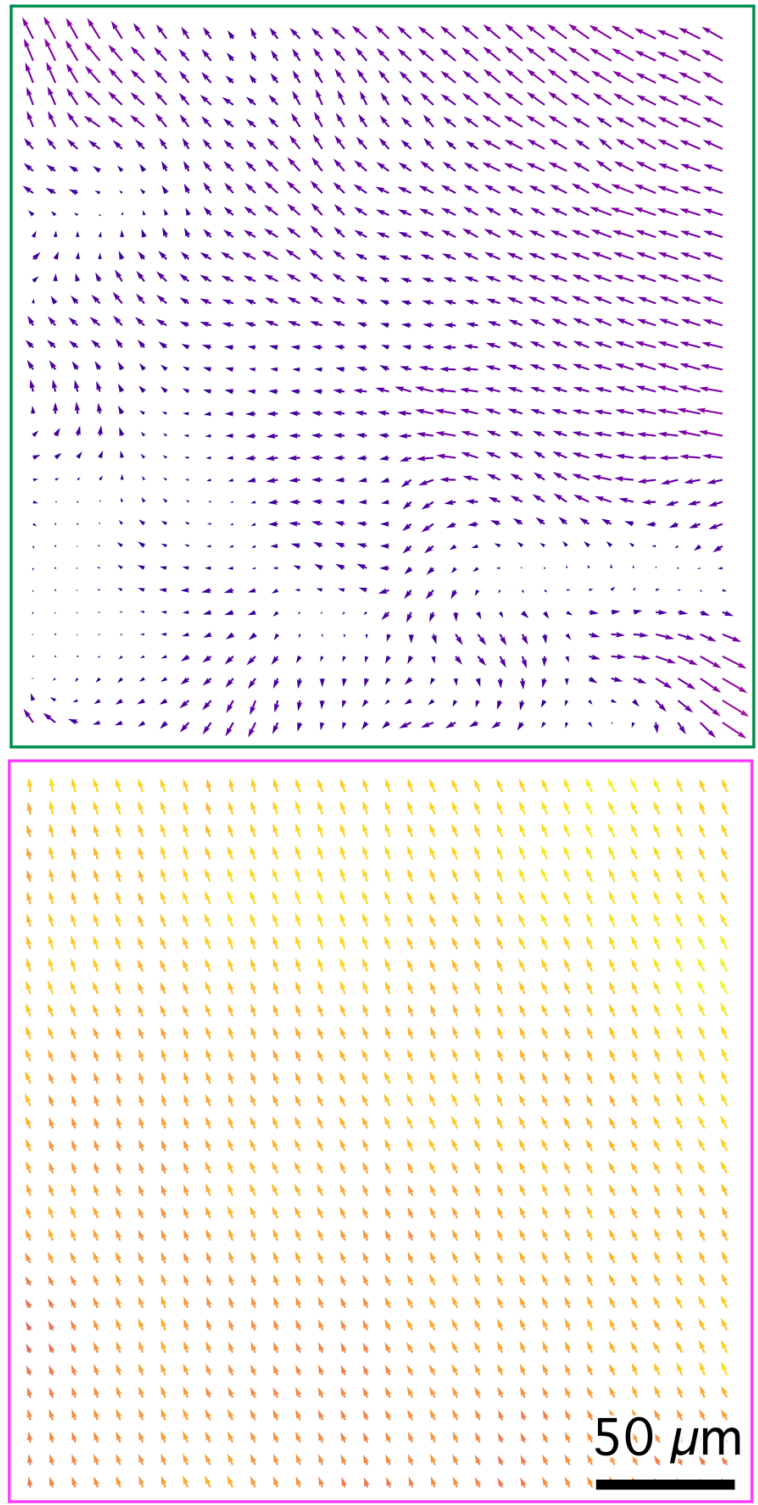
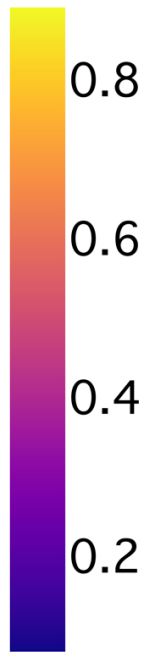

Figure 3H.ii
RDS

speed $v$
($\mu m\ s^{-1}$)

**Figure S2. Velocity vector flow field RDS for the circled data in Figure 3H in the main text.** The full resolution flow field RDS generated with BARCODE from which the down-sampled flow fields shown in Figure 3H.ii are generated. The arrow color and size indicate the speed. The arrow color is normalized relative to both flow fields, whereas the arrow size is scaled relative to the single field it represents. The direction in which the arrow points indicates the velocity direction at each point.

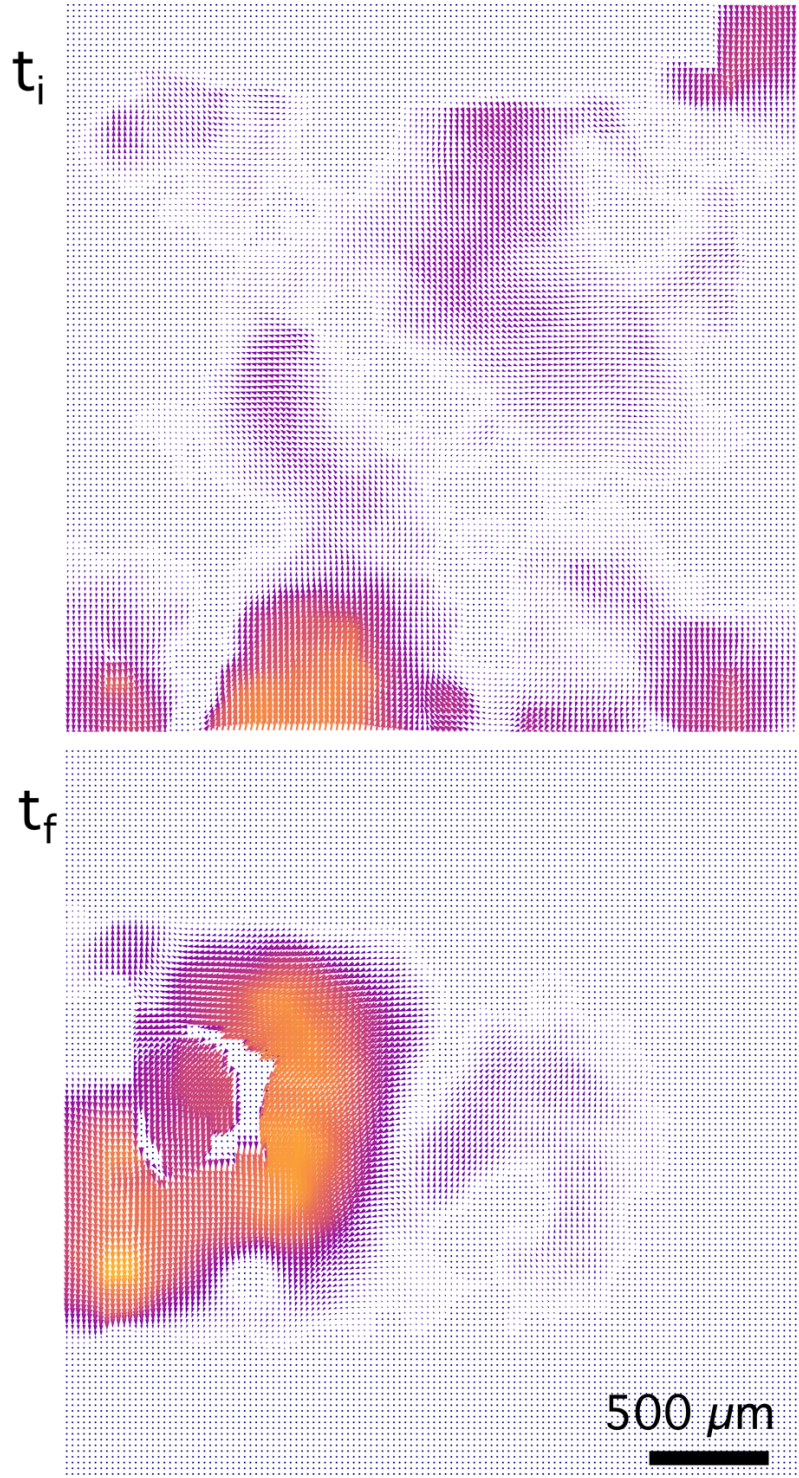
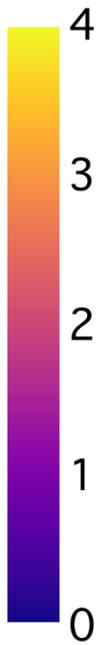

**Figure S3. Velocity vector flow field RDS for the circled iii data in Figure 4D in the main text.** The full resolution flow field RDS generated with BARCODE from which the down-sampled flow fields shown in Figure 4D are generated. The arrow color and size indicate the speed. The arrow color is normalized relative to both flow fields, whereas the arrow size is scaled relative to the single field it represents. The direction in which the arrow points indicates the velocity direction at each point. In regions of very high speed the arrows overlap to a saturating limit so the direction is not discernible at the full resolution.

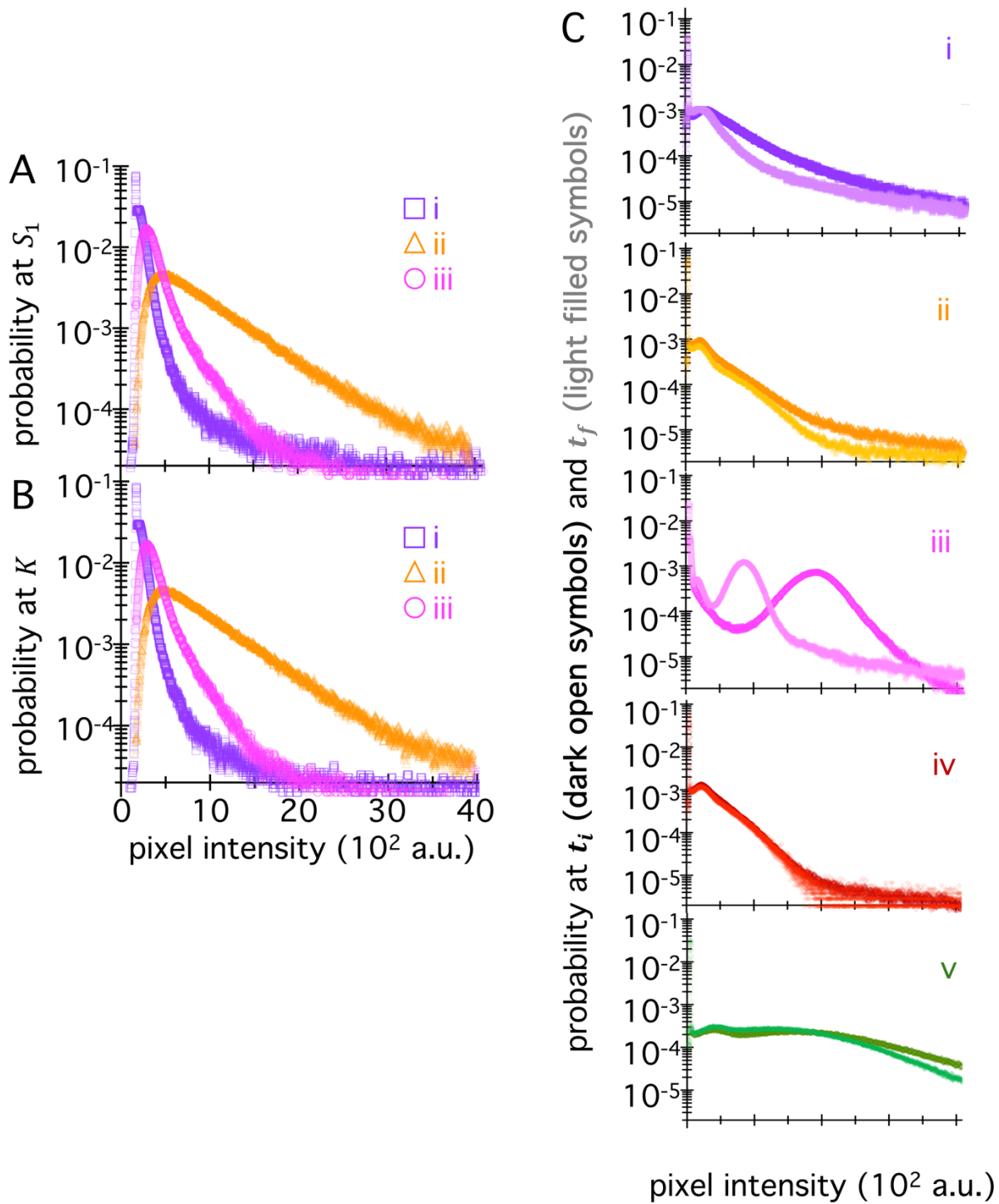

**Figure S4. Intensity probability distributions for the enumerated data points in Figure 4L (left) and 4F (right) in the main text. (A,B)** The distribution for the frame that has the (A) maximum median skewness and (B) maximum kurtosis for the circled data in Figure 4L in the main text labeled as i (purple squares), ii (orange triangles), iii (magenta circles). **(C)** Intensity probability distributions for the initial (dark open symbols) and final (light filled symbols) frames for the data points in Figure 4F that are circled (i-iii) or boxed-in (iv, v).

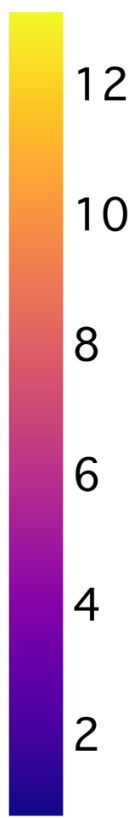
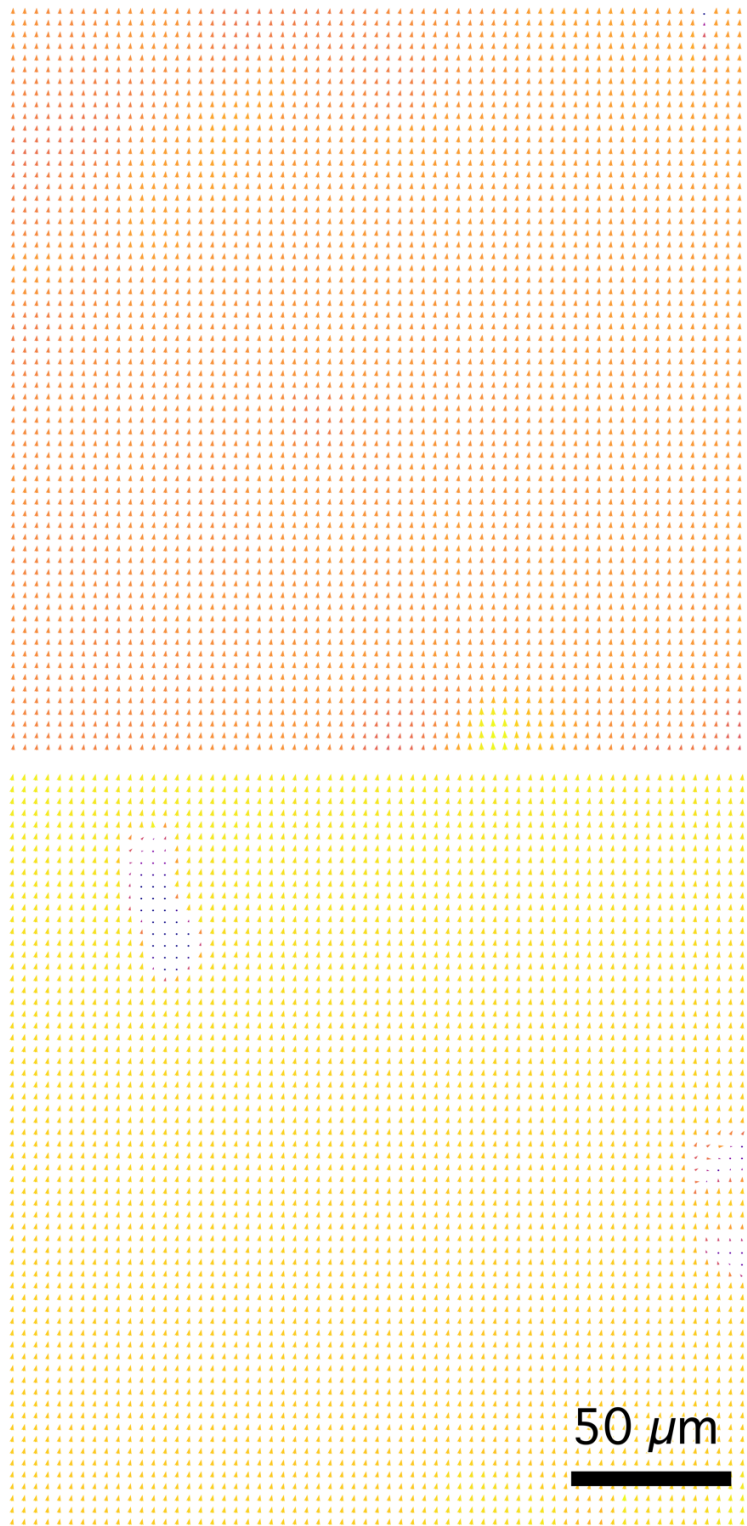

**Figure S5. Velocity vector flow field RDS for the circled i data in Figure 4J in the main text.** The full resolution flow field RDS generated with BARCODE from which the down-sampled flow fields shown in Figure 4J are generated. The arrow color and size indicate the speed. The arrow color is normalized relative to both flow fields, whereas the arrow size is scaled relative to the single field it represents. The direction in which the arrow points indicates the velocity direction at each point.

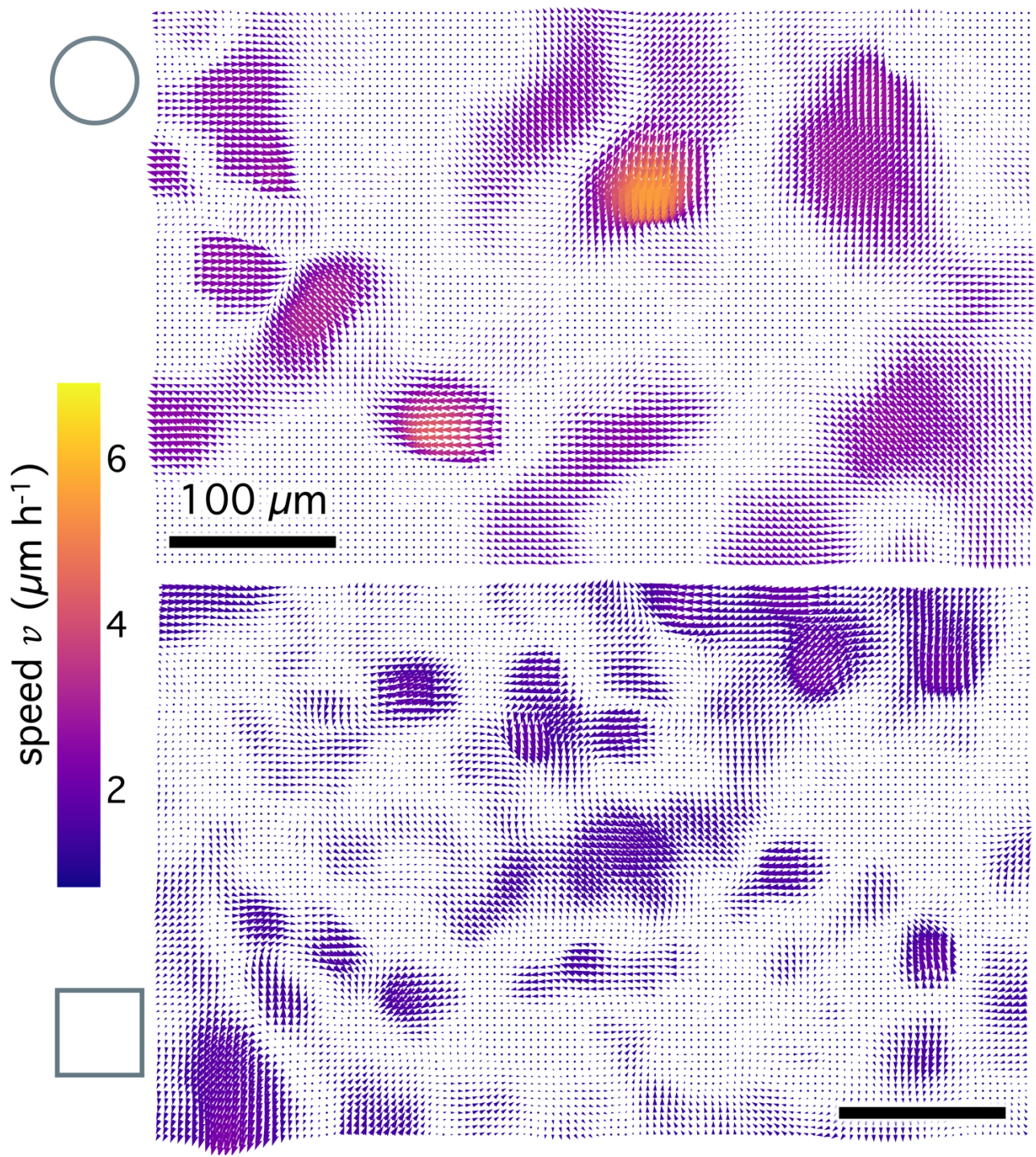

**Figure S6. Velocity vector flow field RDS for the circled data in Figure 5C in the main text.** The full resolution flow field RDS generated with BARCODE from which the down-sampled flow fields shown in Figure 5C are generated. The arrow color and size indicate the speed. The arrow color is normalized relative to both flow fields, whereas the arrow size is scaled relative to the single field it represents. The direction in which the arrow points indicates the velocity direction at each point.

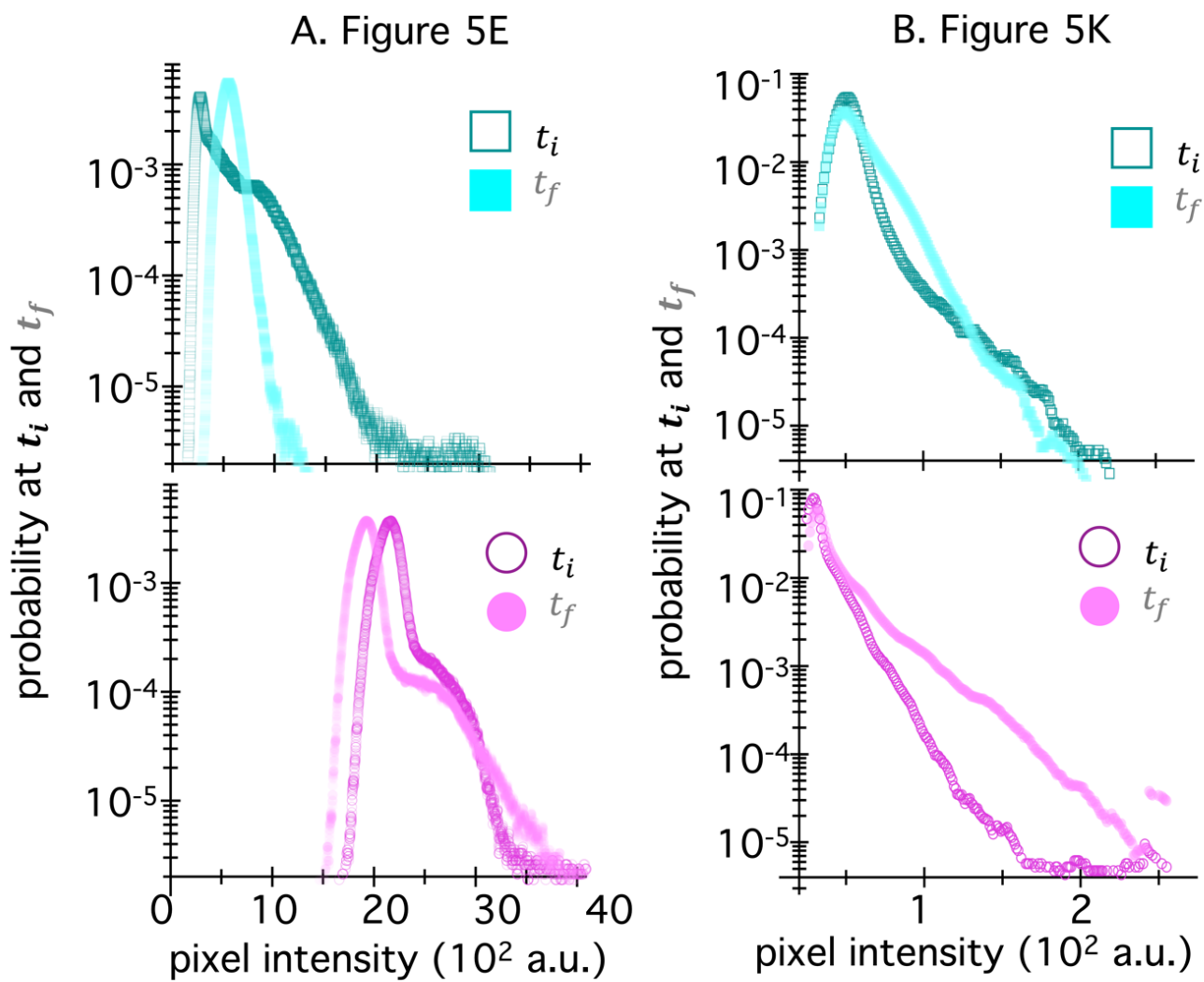

**Figure S7. Intensity probability distributions for the circled data points in Figure 5E (left) and 5K (right) in the main text.** Intensity probability distributions for the initial (dark open symbols) and final (light filled symbols) frames for the circled data points in (A) Figure 5E and (B) Figure 5K. Top plots correspond to the (A) cytoplasm or (B) phase contrast channel data points and bottom plots are for the circled nucleus channel data points.

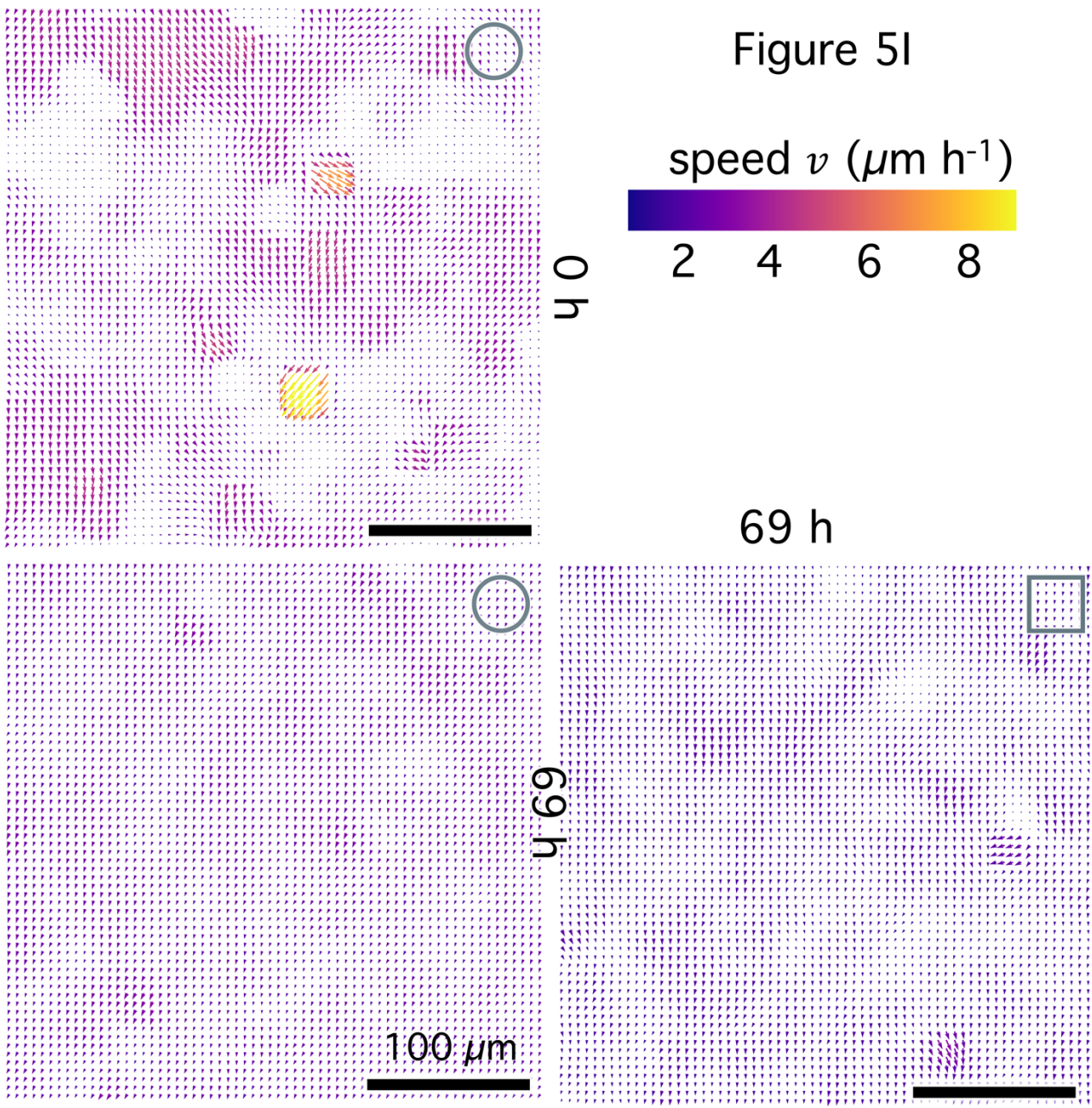

**Figure S8. Velocity vector flow field RDS for the circled data in Figure 5I in the main text.** The full resolution flow field RDS generated with BARCODE from which the down-sampled flow fields shown in Figure 5I are generated. The arrow color and size indicate the speed. The arrow color is normalized relative to both flow fields, whereas the arrow size is scaled relative to the single field it represents. The direction in which the arrow points indicates the velocity direction at each point.

11